\documentclass[structabstract]{aa}  
\usepackage{graphicx}
\usepackage{txfonts}
\usepackage{natbib}

\begin{document}
   \title{Pulse phase and precession phase resolved spectroscopy of \\
   Her X-1: studying a representative Main-On with RXTE}

   \subtitle{}

   \author{D.~Vasco \inst{1}, R.~Staubert\inst{1}, D.~Klochkov\inst{1}, A. Santangelo\inst{1},
          N. Shakura\inst{2}, 
                   \and
          K.~Postnov\inst{2}}

 \institute{
 \inst{1}  Institut f\"ur Astronomie und Astrophysik, Universit\"at T\"ubingen, Sand 1, D-72076 
 T\"ubingen, Germany \\
 \inst{2}  Sternberg Astronomical Institute, Lomonossov University Moscow, Russia \\
                 \email{vasco@astro.uni-tuebingen.de}
          }

   \date{November 2012}

% \abstract{}{}{}{}{} 
% 5 {} token are mandatory
 
  \abstract
   {}
    {We performed a detailed pulse phase resolved spectroscopy of the accreting binary X-ray pulsar 
    Her X-1 in the energy range 3.5--75\,keV and have established pulse phase profiles for all spectral 
   parameters. For three parameters, the centroid energy of the cyclotron line, the photon index 
   and the flux of the 6.4\,keV iron line, we have studied the variation as a function of 35 d phase.}
   {We analyzed RXTE observations of the Main-On of November 2002 
   using data from the PCA and the HEXTE instruments. Four different time intervals of about $\sim$1\,d 
   duration were selected to provide a good coverage of a complete Main-On. 
   The intervals are centered at  35\,d phases 0.03, 0.10, 0.15, and 0.20, respectively.}
    {All spectral parameters show a strong modulation with pulse phase. While the centroid energy of 
   the cyclotron line follows roughly the shape of the pulse profile, showing higher values close to the 
   peak of the X-ray pulse, both the photon index and the iron line intensity exhibit distinct minima 
   around the peak of the X-ray pulse. With respect to variations of the observed profiles with 
   35\,d phase, we find that there is a clear evolution of the shape of the pulse profiles (flux versus pulse 
   phase), a moderate increase of the maximum cyclotron line energy (found around pulse phase 0.7), but no 
   significant evolution of the shape of the pulse phase profiles of the cyclotron line energy, the spectral 
   power law index or the iron line intensity.}
   {
   The variation of spectral parameters as a function of the pulse phase provides important information 
   about the system: i. the disappearance of the Fe line flux near the highest continuum flux 
   may be an indication of a hollow cone geometry of the accretion structure; ii. The apparent non-dependence 
   of the cyclotron line energy profiles on 35\,d phase provides a new possibility to test the model of free 
   precession of the neutron star, proposed to be responsible for the systematic variations in the pulse profiles.}

   
 \authorrunning{D. Vasco, R. Staubert, D. Klochkov, A. Santangelo}
 \titlerunning{RXTE pulse phase spectroscopy of Her~X-1}
 \maketitle
%
%________________________________________________________________

\section{Introduction}
   
The X-ray source Her X-1 was discovered in 1972 by Uhuru \citep{Tananbaum72}, 
and classified as an accreting X-ray binary. Her X-1 is one of the brightest and most studied persistent 
binary X-ray pulsars.
The distance to the system is $\sim$\,7\,kpc and the masses of the neutron star and its companion are 
approximately 1.5 M$_\odot$ and 2.2 M$_\odot$, respectively \citep{Reynolds97}. 
The X-ray flux shows periodic modulation on several different time-scales:
pulsations with 1.24\,s due to the spin of the neutron star, eclipses due to the 1.7\,d orbital period,
and a 35\,d super-orbital period due to obscuration of the X-ray emitting region by the precessing
accretion disk. This 35 d periodicity 
shows different states: two On-states with the $\sim$11\,d Main-On  (at phase 0\,--\,0.31)
and the $\sim$5\,d Short-On (at phase 0.57\,--\,0.79), separated by
two Off-states
(see e.g. \citealt{Giacconi73,Gerend76,Boynton80,Scott99,Scott00,Klochkov06a}). 
The accretion disk is thought to be tilted (with respect to the orbital plane), warped and
counter-precessing with a somewhat variable period of $\sim$35\,d. 
The onset of the flux (often identified with 35\,d phase 0.0) is
called the Turn-On (TO) and corresponds 
to the transition from the Off-state to the Main-On. At this time, the outer rim 
of the disk clears the view to the X-ray emitting region close to the polar caps on the surface of the neutron 
star, leading to an increase in flux. Viceversa, the decline of the
flux towards the end of the Main-On is identified with the inner edge of the accretion disk blocking our view to the X-ray
emitting regions. A similar cycle is responsible for the Short-On with a maximum flux of roughly
one third that of the Main-On.

Her X-1 was also the first accreting X-ray pulsar for which a cyclotron line in the X-ray spectrum has 
been discovered \citep{Truemper78}. This absorption-like feature, now referred to as  Cyclotron 
Resonance Scattering Feature (CRSF), is observed around 40\,keV and allows to estimate the neutron 
star's magnetic field. Applying the formula B$_{12}$ = (1+z) E$_{\rm cyc}$/11.6\,keV (where B$_{12}$ 
is the magnetic field strength in units of $10^{12}$\,Gauss, $z$ is the gravitational redshift and E$_{\rm cyc}$
is the centroid energy of the cyclotron line), the first direct measurement of the magnetic field of a neutron
star was achieved ($\sim3 \times 10^{12}$\,Gauss for Her~X-1). The source shows a positive correlation 
between the centroid energy of the cyclotron line and the bolometric luminosity \citep{Staubert07,Vasco11}. 

Another observational feature of Her X-1 is the evolution of the pulse profile as function of energy and time,
or better: 35\,d phase. For both the Main-On and the Short-On, a systematic variation in the 
shape of the pulse profiles is found. These changes consist of the disappearing of some features in the 
pulse profile with an increase in energy and/or time. These variations repeat systematically on a time scale 
of $\sim$35\,d. These changes have been well known for some time (see e.g. 
\citealt{Deeter98,Scott00,Truemper_etal86}).  Recently, \citet{Staubert_etal10a,Staubert_etal10b, Staubert_etal12} 
on the basis of RXTE observations, gave a quantitative description of these changes by providing a 
template, which can be used to predict the shape of the profile in the 9--13\,keV energy range for any 35\,d 
phase during the Main-On. An extension of this work for other energy ranges and including the Short-On is in preparation.

  % Table 1   __________________________________________________
   \begin{table}[t!]
      \caption[]{Summary of the four intervals of the Main-On of 35\,d cycle no. 323.
      The columns are: the number of the interval, the time interval
      in MJD, the exposure time, the corresponding 35\,d 
      phase interval, and the center of the 35\,d phase interval.}
         \label{observations}
\vspace{-6mm}
     $$ 
         \begin{array}{ccccc}
            \hline
            \noalign{\smallskip}
%          {\rm interval}  &  {\rm MJD}                                                          & 35\,{\rm d}\,{\rm \,phase}    & \mathbf{35\,{\rm d}\,{\rm \,phase}} \\
             {\rm Interval}  &  {\rm limits}                 & \mathbf{\rm exposure}  & {\rm limits}                           &  \mathbf{{\rm center} } \\
                                   &   {\rm MJD}                 &  \mathbf{\rm [ks]}          & 35\,{\rm d}\,{\rm \,phase}     &  \mathbf{\rm 35\,{\rm d}\,{\rm \,phase} } \\
            %\noalign{\smallskip}
            \hline
            \noalign{\smallskip}
            1                    & 52595.060-52596.743  &  \mathbf{\rm 25}                   & 0.007\,-\,0.056                      & \mathbf{\rm 0.03}  \\
            2                    & 52597.959-52598.719  &  \mathbf{\rm 25}                   & 0.090\,-\,0.112                      & \mathbf{\rm 0.10} \\
            3                    & 52599.603-52600.726  &  \mathbf{\rm 25}                  & 0.137\,-\,0.161                      & \mathbf{\rm 0.15} \\
            4                    & 52601.316-52602.406  &  \mathbf{\rm 28}                   & 0.186\,-\,0.217                      & \mathbf{\rm 0.20} \\
            %5   & 52603.027\,-\,52604.032  & 0.25    \\
            \noalign{\smallskip}
            \hline
         \end{array}
     $$ 
%\begin{list}{}{}
%\item[$^{\mathrm{a}}$] This is footnote a
%\end{list}
   \end{table}

\citet{Truemper_etal86}, on the basis of observations by Exosat, had proposed that
free precession of the neutron star is responsible for those pulse profile changes, due to the 
variation of the angle of the line of sight to the X-ray emitting regions on the surface of the neutron star.
For Her X-1, the dependence of the spectral parameters on pulse phase have been observed by several 
missions. Early examples are \cite{Pravdo78} with OSO 8, \cite{Pravdo79} and \cite{Soong90b} 
with the A-4 experiment of HEAO-1 and \cite{McCray82} with HEAO 2 the Einstein Observatory,  \cite{Voges82} with the MPE/AIT ballon experiment, \cite{Kahabka_87} 
with Exosat, \cite{Kunz96} with Mir-HEXE.  
Later examples are  \cite{Endo00} with ASCA, \cite{Zane01} with \emph{XMM-Newton}, 
\cite{Lutovinov00} with the ART-P telescope on board of GRANADA and 
\cite{Klochkov08} with INTEGRAL. 
 
 In this work we present a deep investigation of the variation of key spectral parameters
 as a function of pulse phase with the finest resolution so far and, for the first time, a discussion of a
 possible evolution with 35\,d phase. Changes of the spectral parameters with pulse phase are quite 
 common among accreting X-ray pulsars and are generally attributed to a change in the viewing 
 angle of the accretion region (see e.g., \citealt{Kreykenbohm04} and references therein). Here, we focus 
 on the variation of the centroid energy E$_{\rm cyc}$ of the CRSF
 around 40\,keV, the photon index $\Gamma$ and the iron line characteristics. We also address the 
 question whether our data support the free precession model which has been suggested to explain the
 variation in shape of the pulse profile.

\section{Observations and data analysis}

\begin{figure}[t!]
  \centering
      \includegraphics[width=0.50\textwidth]{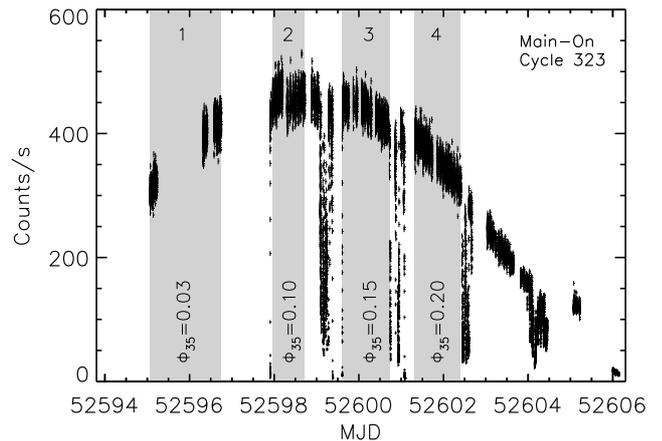}  
  \caption{PCA light curve of the Main-On of cycle 323. The grey zones represent the four intervals 
  used for this analysis. For each interval the centered 35\,d\,phase is given. Details of the time 
  and 35\,d\,phase intervals are listed in Table \ref{observations}}.
   \label{obsint}
\end{figure}

To carry out this analysis we used observations of Her X-1 of the \emph{Rossi} X-ray Timing Explorer 
(RXTE) performed in November 2002. These observations correspond to cycle number 323 (cycle numbering 
according to Staubert et al. 2009b) centered at MJD $\sim$52599.32 with an observed Turn-On of 
MJD~52594.80. The data of this cycle provide the best coverage of any Main-On of Her~X-1 
observed with RXTE. Within this cycle, we sum up observations from four selected 
time intervals at different 35\,d phases in order to have a good coverage of an entire Main-On, 
avoiding dips and eclipses. All intervals have a duration of about one day (details are given in 
Table \ref{observations}). Figure \ref{obsint} shows the four intervals selected for this analysis.

%Fig. 1

%__________________________________________________________________

%Fig. 2
\begin{figure}[t!]
  \centering
      \includegraphics[width=0.50\textwidth]{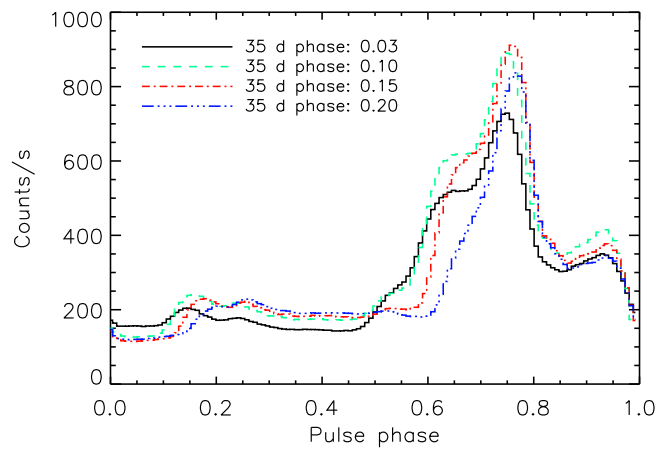}  
  \caption{Mean 9-13\,keV pulse profiles of the four time intervals at different 35\,d phases (see Table \ref{observations}):
  Interval 1 in black (solid line), Interval 2 in green (dashed line),
  Interval 3 in red (dash dotted line) and Interval 4 in blue (dash
  dot dot dotted line). For general 35\,d
  dependence see  \citet{Staubert_etal10a,Staubert_etal10b}. As timing reference for pulse phase zero 
  we use the ``sharp-edge'' at the trailing edge of the right shoulder of  the main pulse which leads into 
  an eclipse like minimum (Staubert et al. 2009b). Note that these pulse profiles 
  %(in the 9\,-\,13\,keV energy range) 
  are statistically so accurate that the uncertainties are smaller than the line width.}
   \label{pp}
\end{figure}

For barycentric and binary corrections orbital parameters and the ephemeris of \citet{Staubert09b} were used: 
$P_{\rm orb}$=1.700167287\,s, $\dot{P}_{\rm orb}$\,=\,+(2.8\,$\pm$\,0.2)\,$\times10^{-12}\,s\,s^{-1}$,\,$a\cdot\,\sin\,i$\,=\,13.1831 lt\,-\,s and $T_{\frac{\pi}{2}}$=52599.486440 MJD. 
To align the pulse profiles, a reference time $t_{0}$ = MJD~52594.869900091 and 
a spin period $P_{\rm spin}$ = 1.237761809\,s were used. As  ``pulse phase zero''  the sharp-edge 
feature at the trailing edge of the right shoulder of the main peak was used (see \citealt{Staubert09b}).

%Fig. 3
 \begin{figure*}[th!]
  \centering
     \includegraphics[bb=95 417 533 694,width=8cm]{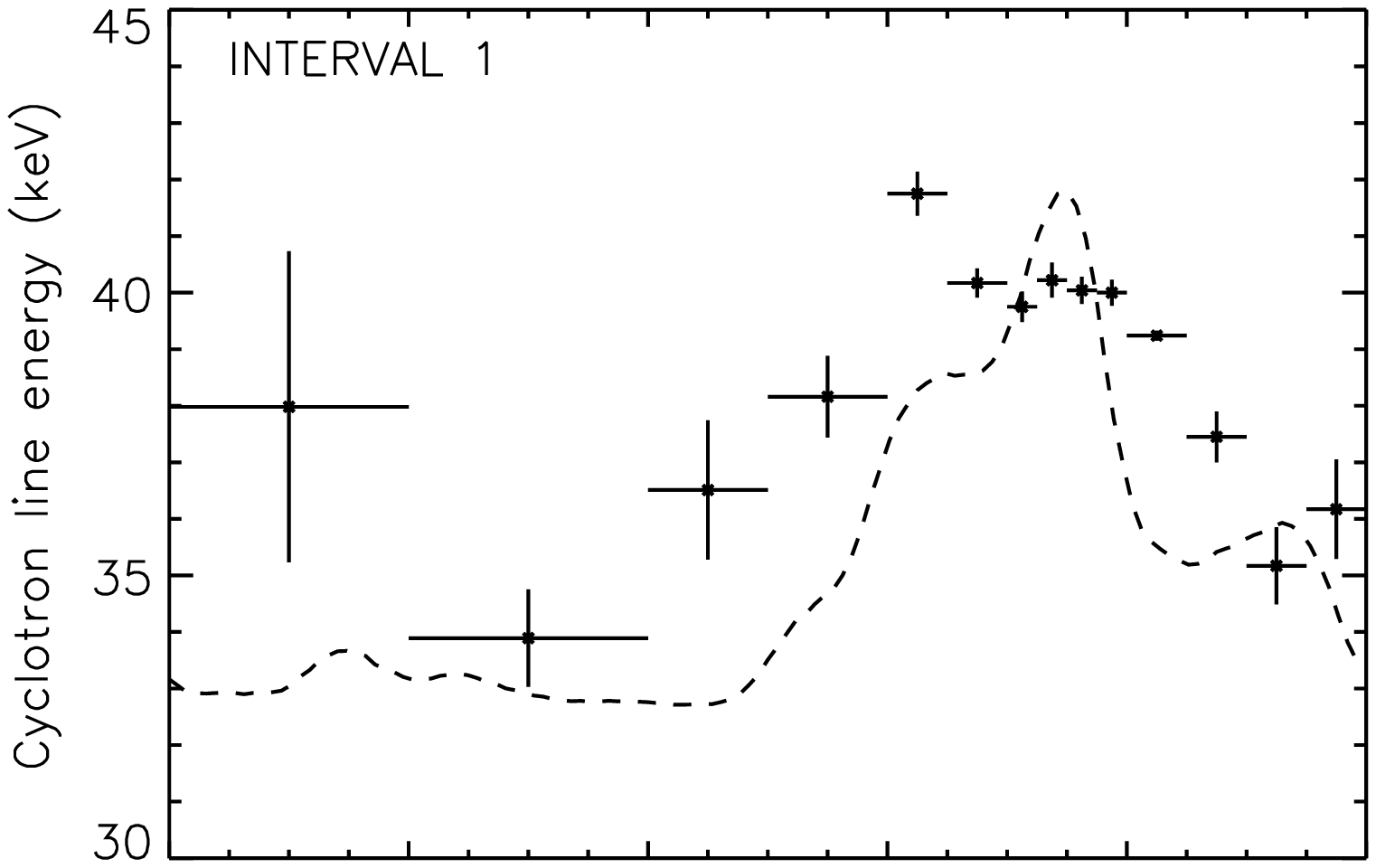} 
     \includegraphics[bb=143 417 533 694,width=7.125cm]{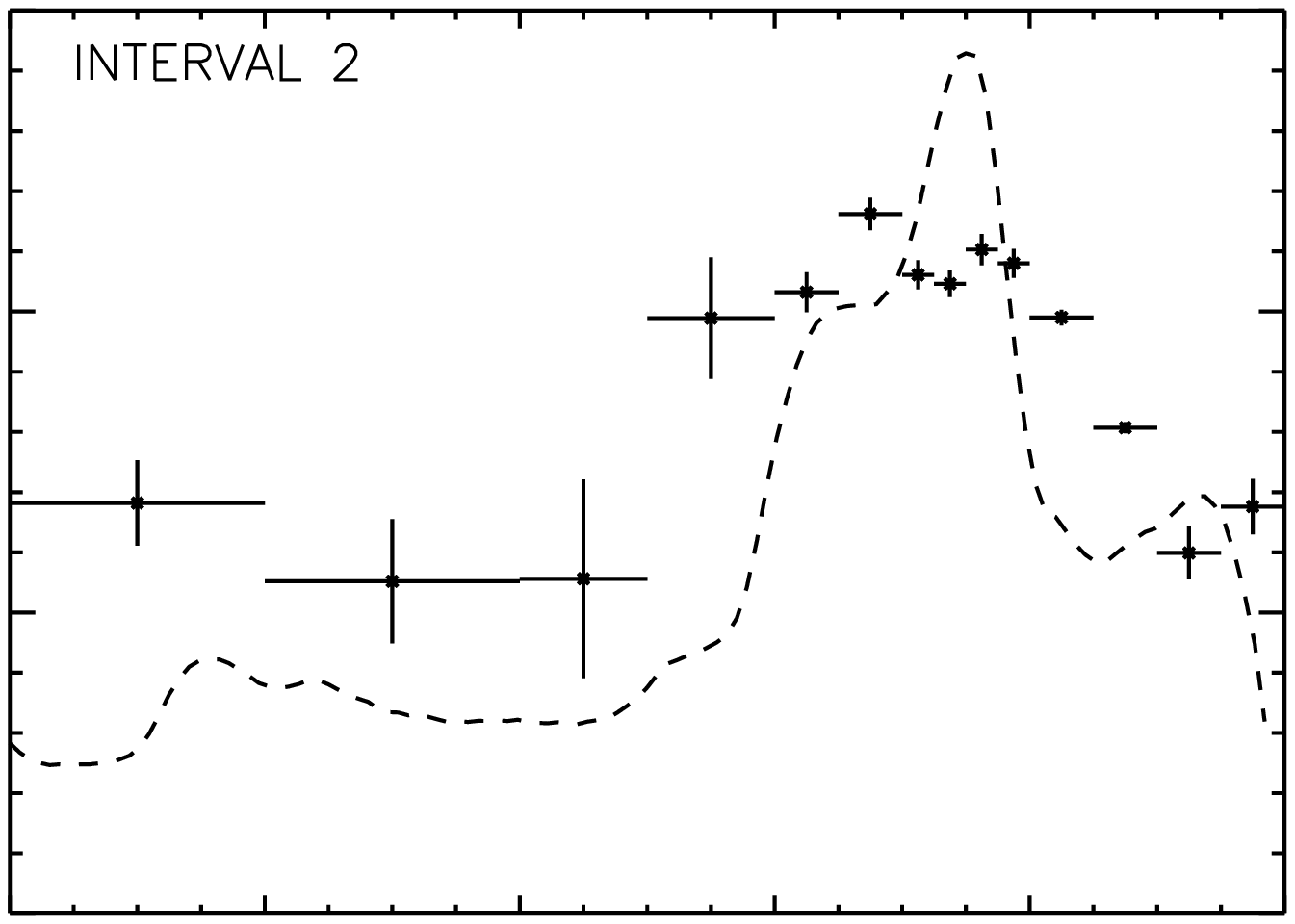}\\
     \includegraphics[bb=95 371 533 694,width=8cm]{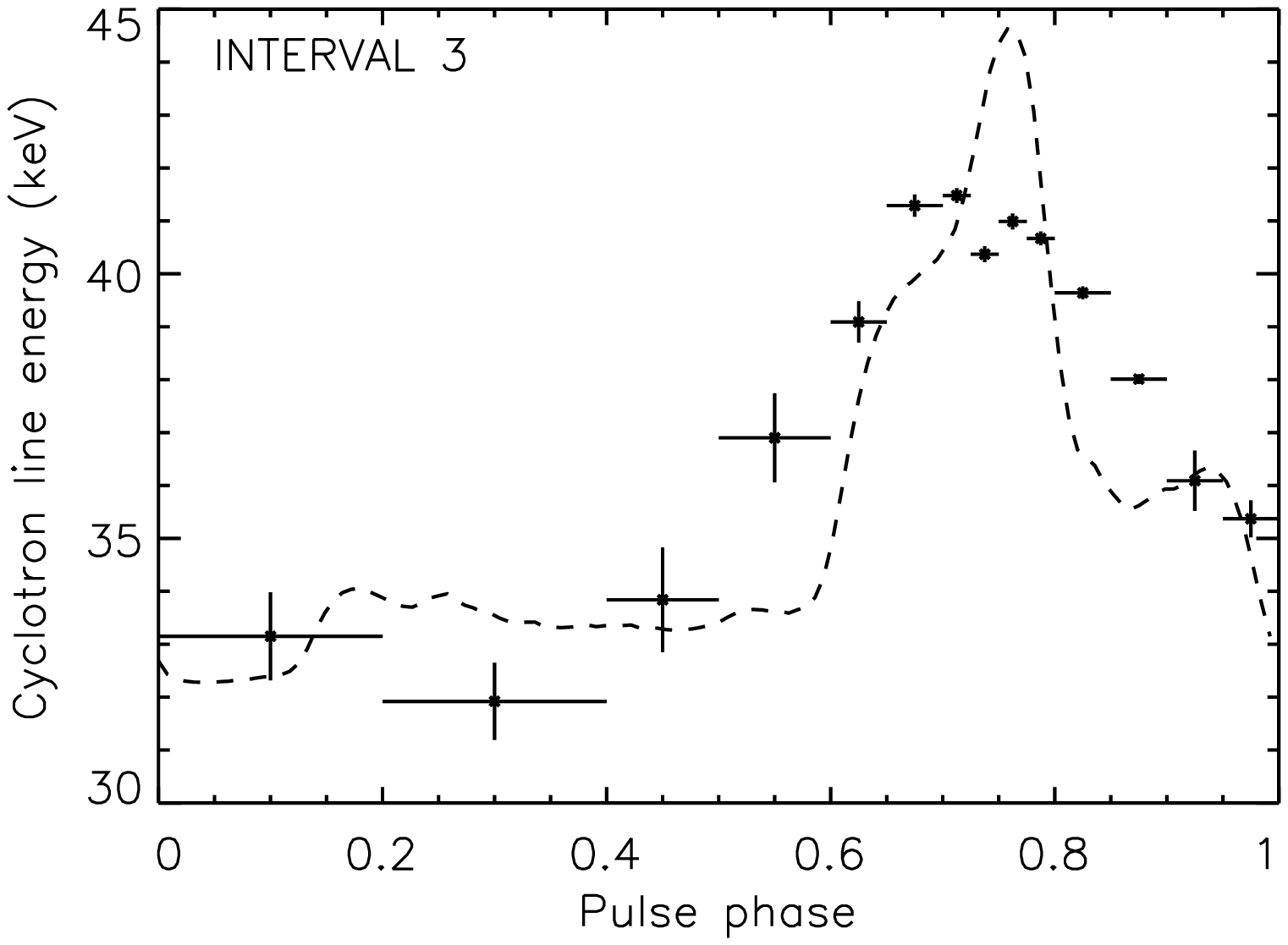}
     \includegraphics[bb=143 371 533 694,clip,width=7.125cm]{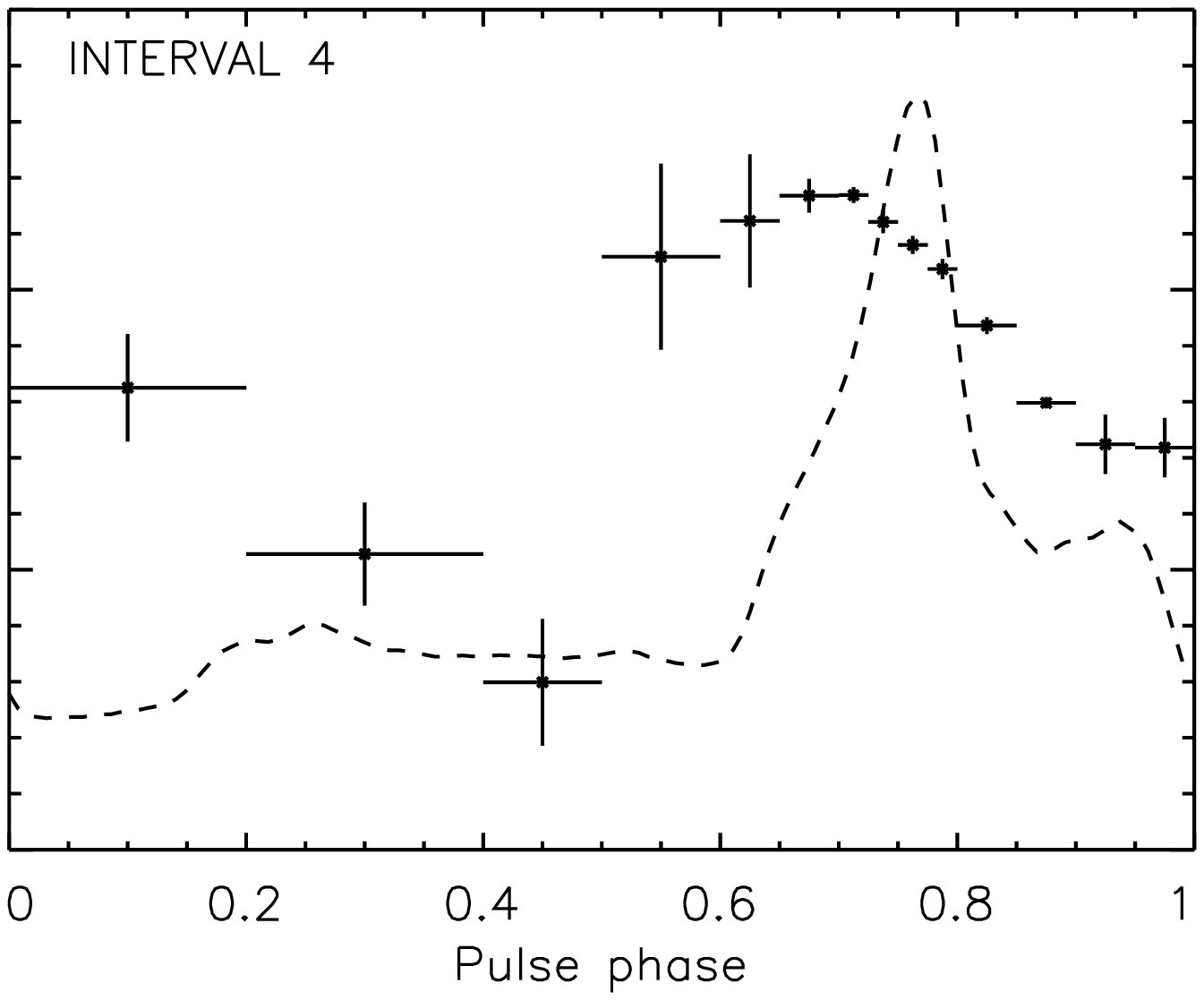}
\caption{Cyclotron line centroid energy profiles as function of the pulse phase for Interval 1 (\textsl{top left}), 
Interval 2 (\textsl{top right}), Interval 3 (\textsl{bottom left}) and Interval 4 (\textsl{bottom right}).}
   \label{cyclll}
\end{figure*}

%Fig. 4
\begin{figure}[b!]
    \includegraphics[width=0.50\textwidth]{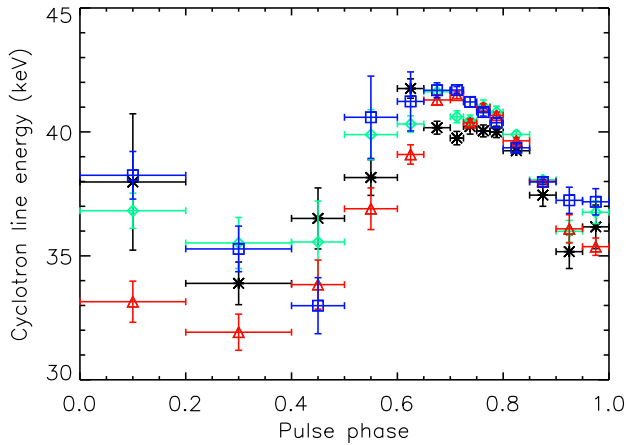}
   \caption{ The cyclotron line centroid energy as function of pulse phase for the four 
   different 35\,d phase intervals (see Table \ref{observations}):
  Interval 1 in black (asterisks), Interval 2 in green (diamonds),
  Interval 3 in red (triangles) and Interval 4 in blue (squares).}
   \label{compar_e}
\end{figure}

The PCA was used in the energy range 3.5\,--\,60\,keV and HEXTE in the energy range 20\,--\,75\,keV. 
The data have been analyzed with XSPEC 12.6.0\footnote{http://heasarc.gsfc.nasa.gov/docs/xanadu/xspec} 
using the \texttt{highecut}\footnote{http://heasarc.gsfc.nasa.gov/xanadu/xspec/manual/XSmodelHighecut.html} 
 spectral model, power law continuum with an exponential cut-off. The differential photon flux
 f(E) is described by:

%Fig. 5
\begin{figure*}[t!]
\centering
     \includegraphics[bb=95 417 533 694,width=8cm]{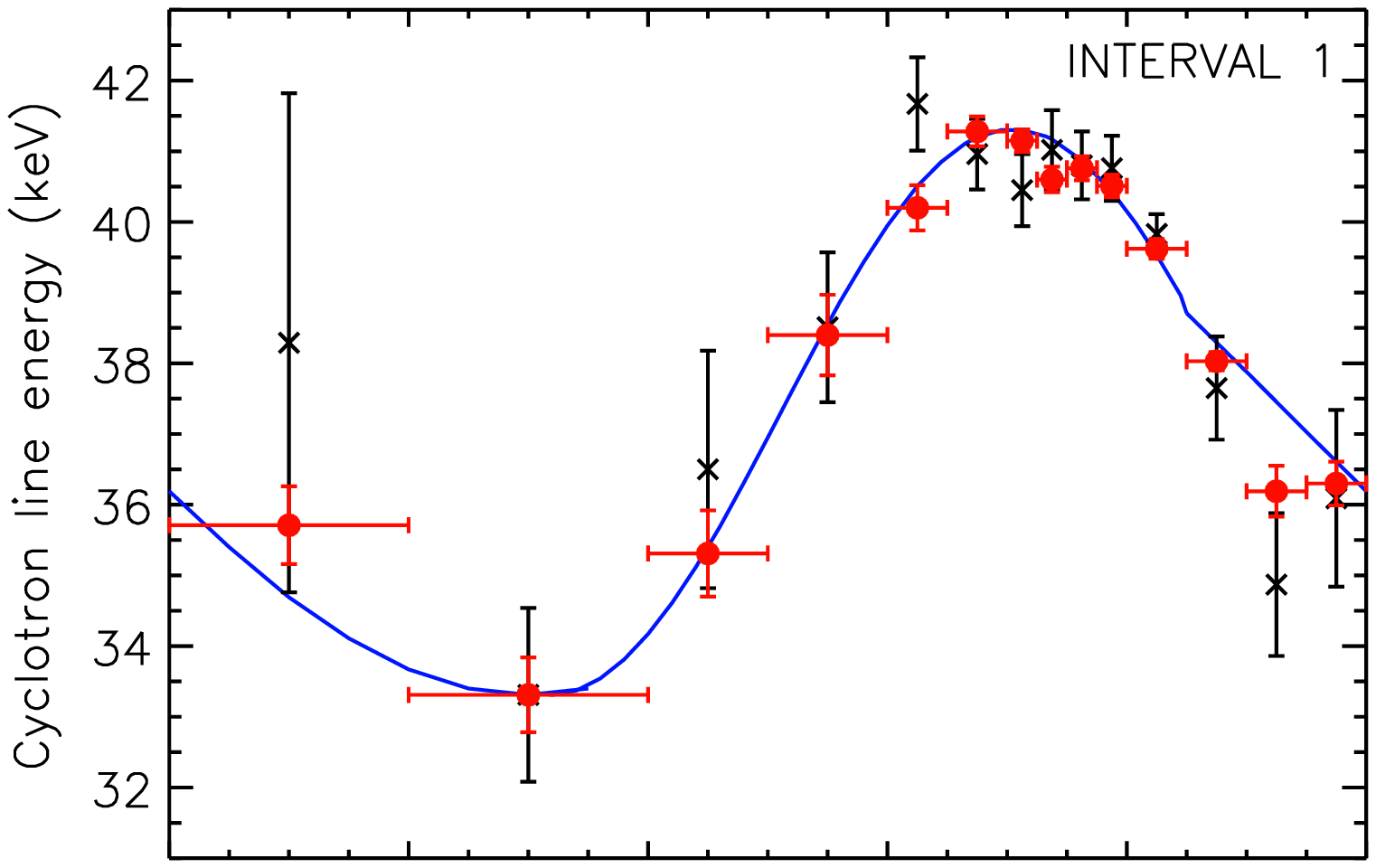} 
     \includegraphics[bb=143 417 533 694,width=7.125cm]{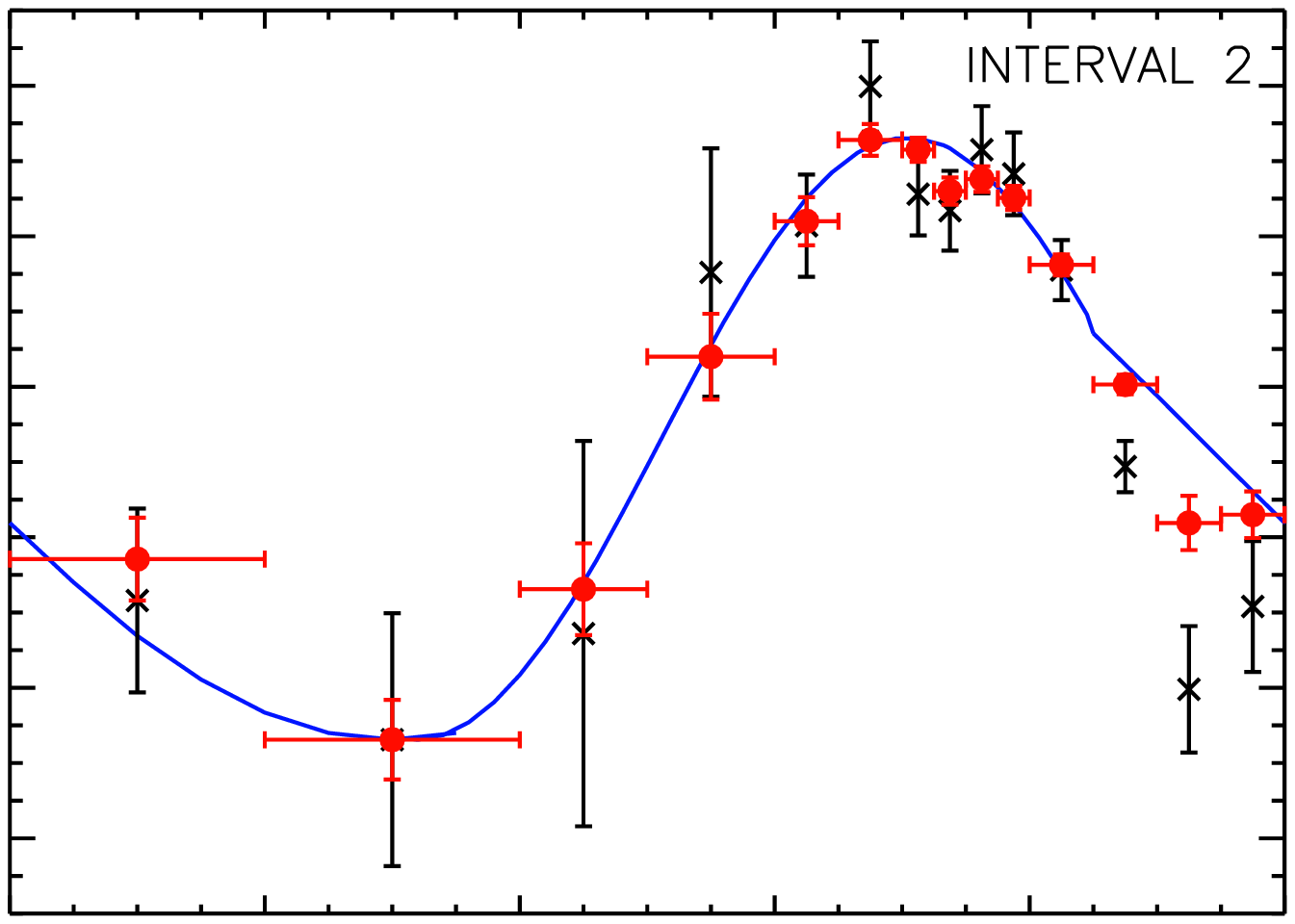}\\
     \includegraphics[bb=95 371 533 694,width=8cm]{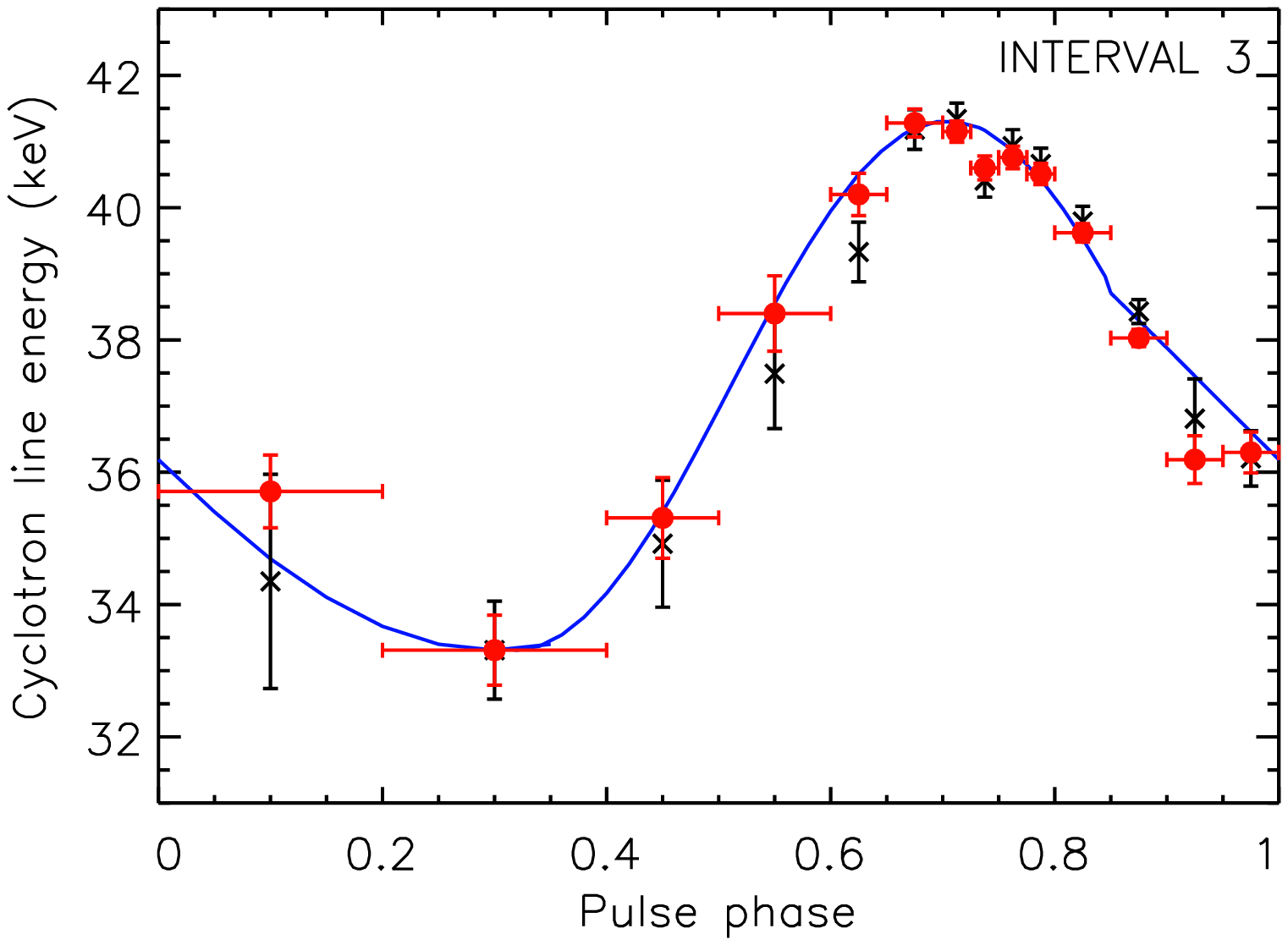}
     \includegraphics[bb=143 371 533 694,clip,width=7.125cm]{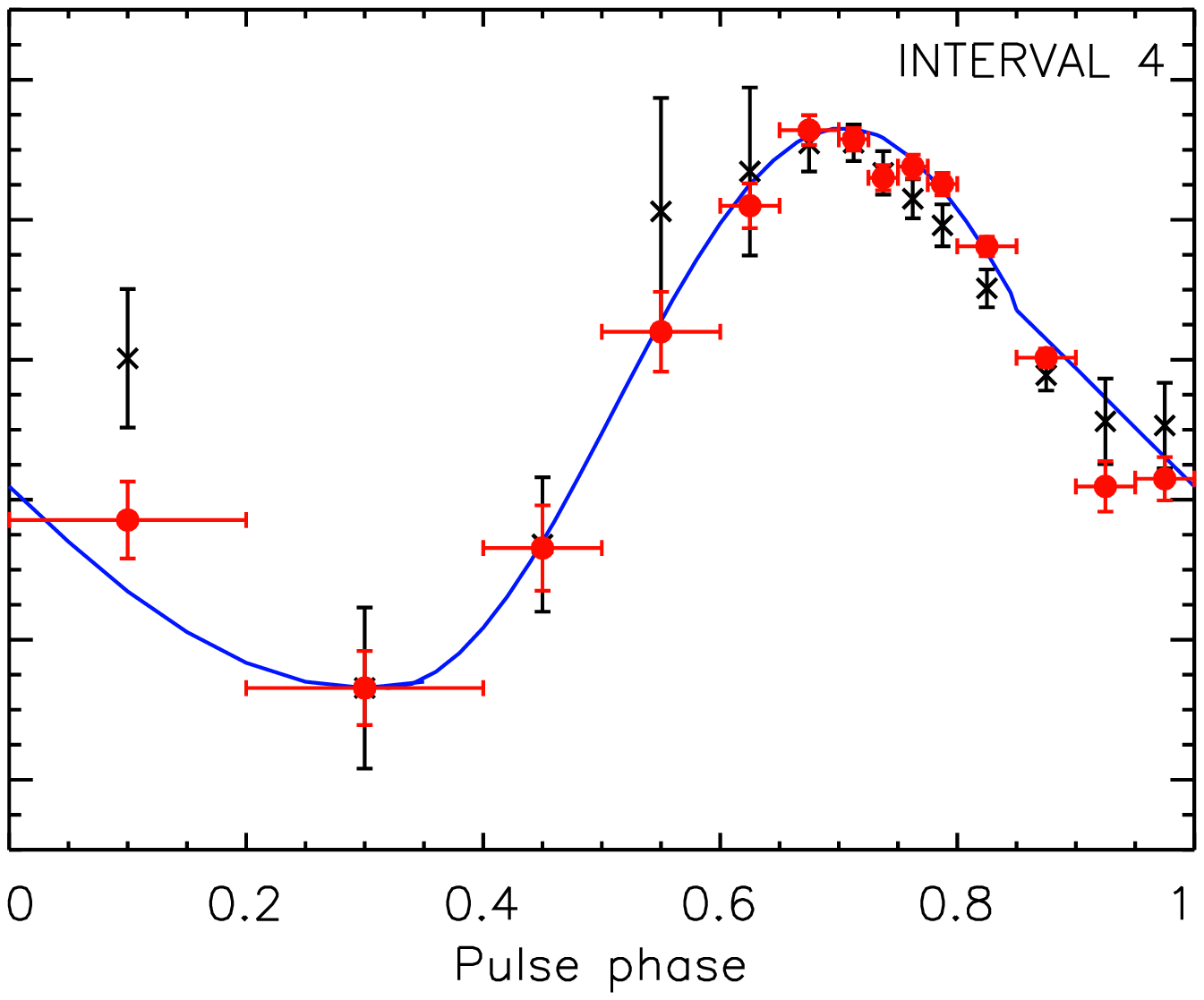}
  % \caption[Photon-index profiles]{\small{The cyclotron line profiles
\caption{\small{The cyclotron line profiles of Fig.~3 for the four 35\,d intervals (black crosses): Interval 1
       (\textsl{top left}), Interval 2 (\textsl{top right}), Interval 3
       (\textsl{bottom left}) and Interval 4 (\textsl{bottom right}), normalized to a common amplitude. 
       The big red filled dots represent the weighted mean values of all four intervals. The solid blue line 
       represents a best fit function (a combination of two cosine components). We conclude that, within the given 
       statistics, there is no evidence for any change in shape of the profiles with 35\,d phase.}}
   \label{normalized2}
\end{figure*}

\begin{equation}
f(E) = A 
\left\{
\begin{array}{rl}
E^{-\Gamma} ~~~~~~~~~~~~~~~~~~~~~, & \mbox{if } E \leq E_{\rm cut} \\ 
E^{-\Gamma}\cdot \mbox{exp} \left( \frac{E-E_{\rm cut}}{E_{\rm fold}}\right), & \mbox{if }  E > E_{\rm cut}
\end{array}
\right.
\end{equation}

 where  $\Gamma$ is the photon index, E$_{\rm cut}$ is the cut-off energy and E$_{\rm fold}$ is
 the e-folding energy. In addition, two line features are 
 included in the model: a multiplicative \textsl{Gaussian} optical depth for the Cyclotron Resonant 
 Scattering Feature (CRSF) and an additive \textsl{Gaussian} emission line for the iron fluorescence 
 feature around 6.5\,keV. This model is the same one as used in earlier analyses (see e.g., \citealt{Coburn02} 
 and \citealt{Staubert07}). We have consistently found that the \texttt{highecut} spectral model describes the
 spectrum of Her~X-1 best. The discontinuity of this model at the cut-off energy does not constitute a problem and
 we have verified that no systematic uncertainties are introduced in estimating the spectral parameters.
 We used the PCA data up to 60\,keV, which has become feasible by new response matrices (PCA response 
 v11.7, 2009 May 11). These high energy data can definitively improve the photon statistics at energies around 
 and beyond the cyclotron line feature. The same choice has been used in previous analyses 
 (see e.g. \citealt{Ferrigno11}) and it was confirmed  by
 \citet{Rothschild11} with RXTE observations of Cen A. 
 The \texttt{recorn}\footnote{http://heasarc.gsfc.nasa.gov/xanadu/xspec/manual/
 XSmodelRecorn.html}
 spectral component was also added to the model to normalize the background. 
 After we had verified, that the energy and width of the iron~K line is constant (6.45\,keV and 0.5 keV, respctively),
 we fixed those values in order to minimize the number of free parameters.
 GoodXenon\footnote{http://heasarc.gsfc.nasa.gov/docs/xte/abc/
 pca$\_$issues.html$\#$configs$\_$modes}
 observational modes of the PCA for high resolution analysis were used to exctract light curves and spectra.
 An additional systematic uncertainty of 1\% was added in using the PCA data. To analyze the data, 
 version v6.11 of the HEASOFT software\footnote{http://heasarc.gsfc.nasa.gov/ftools/} and background
 estimation files Sky$\_$VLE dated 2005 November 28 were used. We have investigated whether there 
 are correlations between the fit parameters: no dependence of any single parameter on the others has 
 been found. All spectral results stated below have been obtained through spectral fits with the assumption of
negligible absorption due to neutral hydrogen, both intrinsic to the source and interstellar (which is at
5.1 $\times$~$10^{19}$ cm$^{-2}$, \citealt{DalFiume_etal98}). The introduction of absorption as a free 
parameter does not lead to a reduction in $\chi^{2}$.

For the pulse phase resolved analysis, we have extracted the spectra in 14 pulse phase bins 
over the 1.24\,s pulse, the widths of which were chosen to have roughly identical photon-statistics in each
spectrum. The centers of the 14 phase bins are at pulse phases: 0.1, 0.3, 0.45, 0.55, 0.625, 0.675, 0.7125, 
0.7375, 0.7625, 0.7875, 0.8250, 0.8750, 0.9250, and 0.9750. For the cyclotron line energy 
E$_{\rm cyc}$ this is the finest resolution in pulse phase ever achieved (the four smallest bins
around the peak of the pulse have a width of 1/80 of a phase).

With the spectral function described above, good fits with reduced $\chi^{2}$ between 0.9 and 1.2 are generally
achieved. However, splitting up the data into small pulse phase intervals reveals that larger $\chi^{2}$ values
appear: up to $\sim$12 for the pulse phase range 0.775--0.85, and 1.4--2.8 for the pulse phase range 0.5--0.775.
The main reason for this is the existence of the so called "10 keV feature" (see the discussion in 3.5). Modeling
this feature by an extra Gaussian is generally successful and brings the $\chi^{2}$ down to acceptable
values (in a few cases $\chi^{2}$ is still found around 1.3, which is most likely due to an imperfect modeling
by  a simple Gaussian).

\section{Pulse phase variability}

We concentrate on the detailed study of the variation of the following spectral parameters as function 
of pulse phase: the centroid energy E$_{\rm cyc}$ of the cyclotron line, the photon index $\Gamma$, 
and the intensity of the iron line at 6.4\,keV. The phase profiles of these spectral parameters are 
compared with the shape of the flux profiles (the "pulse profiles")
for the four different 35\,d phase intervals. In Figure \ref{pp} the 9\,--13\,keV pulse 
profiles of the four intervals are shown. 
Note that the pulse profiles are of such high statistical quality
that the uncertainties of the flux values (in each of the 128 pulse phase bins) are comparable to the width
of the lines chosen to plot the profiles. The profiles are repeated in the following figures (as dashed 
lines) to allow a direct comparison with the profiles of the spectral parameters. For the remaining 
spectral parameters, such as the cut-off energy  E$_{\rm cut}$, the folding energy E$_{\rm fold}$ 
(with reference to the used spectral model - \texttt{highecut}), and the width and depth
of the cyclotron line ($\sigma_{\rm cyc}$ and $\tau_{\rm cyc}$), we will restrict ourselves to profiles of 
the combined data of cycle 323. For these parameters the statistics is such, that no
statements can be made about a dependence on 35\,d phase. In all Figures, the uncertainties 
given are one sigma (68\%) values.

%Fig. 6
\begin{figure*}[t!]
\centering
     \includegraphics[bb=95 417 533 694,width=8cm]{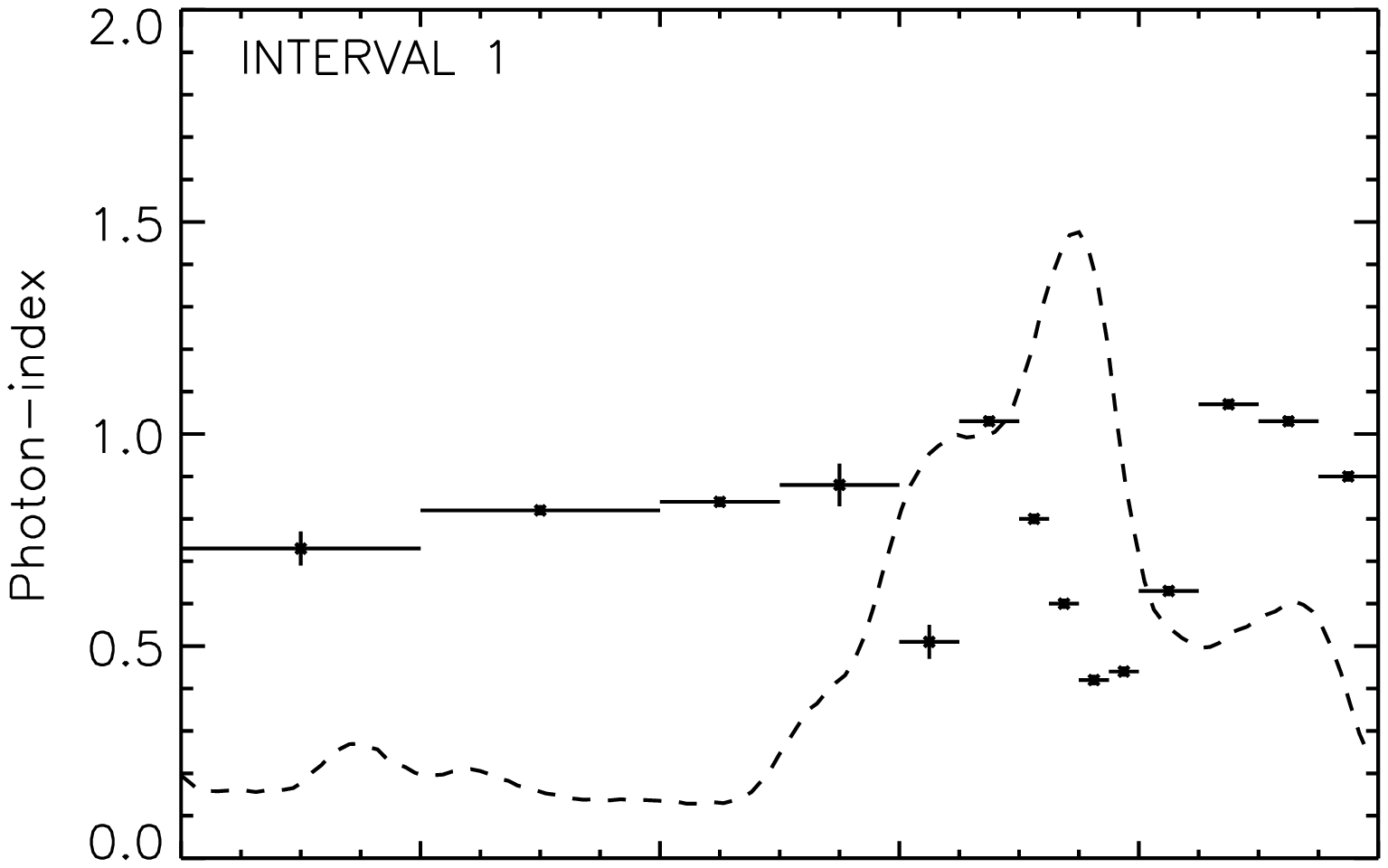} 
     \includegraphics[bb=143 417 533 694,width=7.125cm]{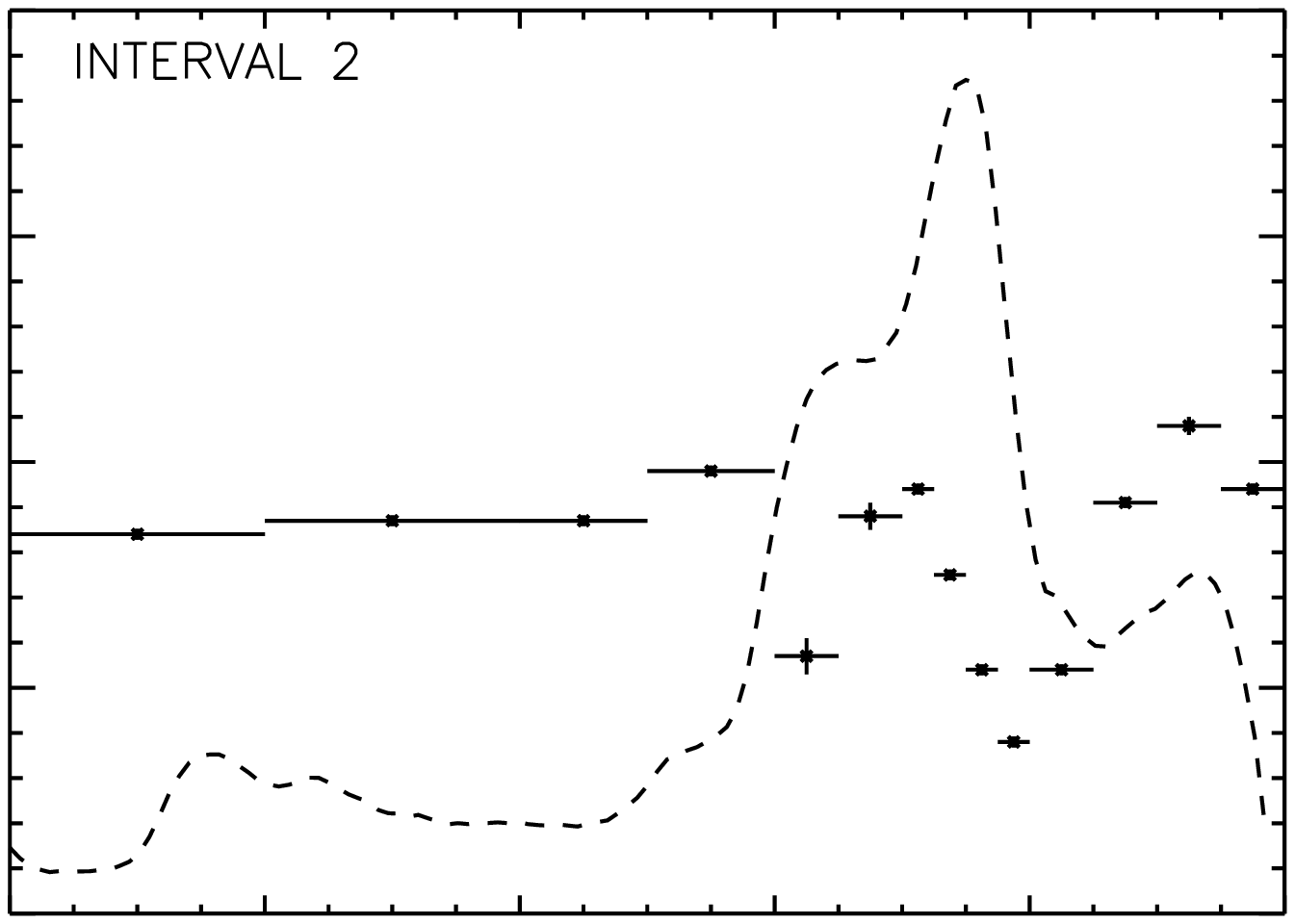}\\
     \includegraphics[bb=95 371 533 694,width=8cm]{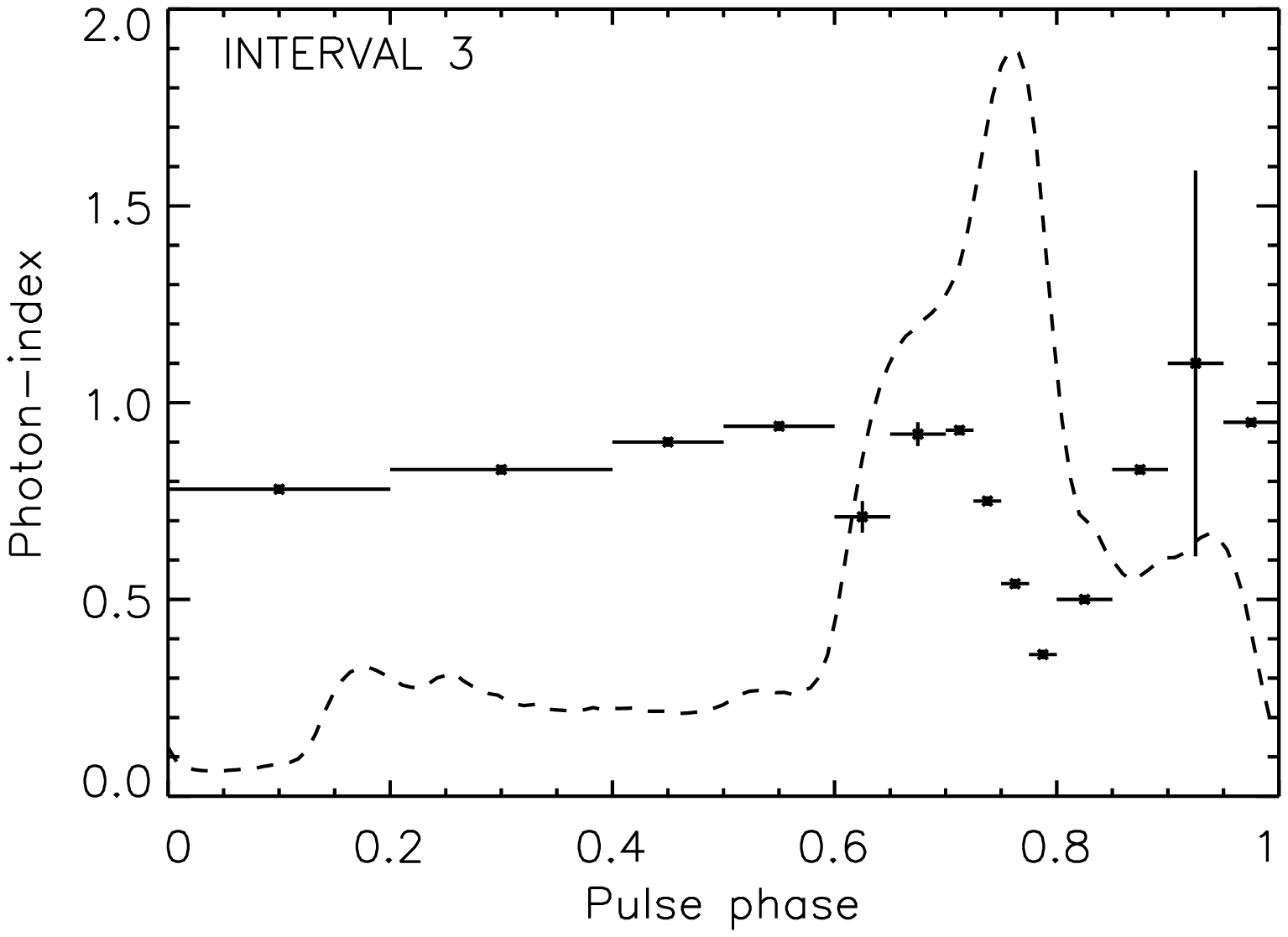}
     \includegraphics[bb=143 371 533 694,clip,width=7.125cm]{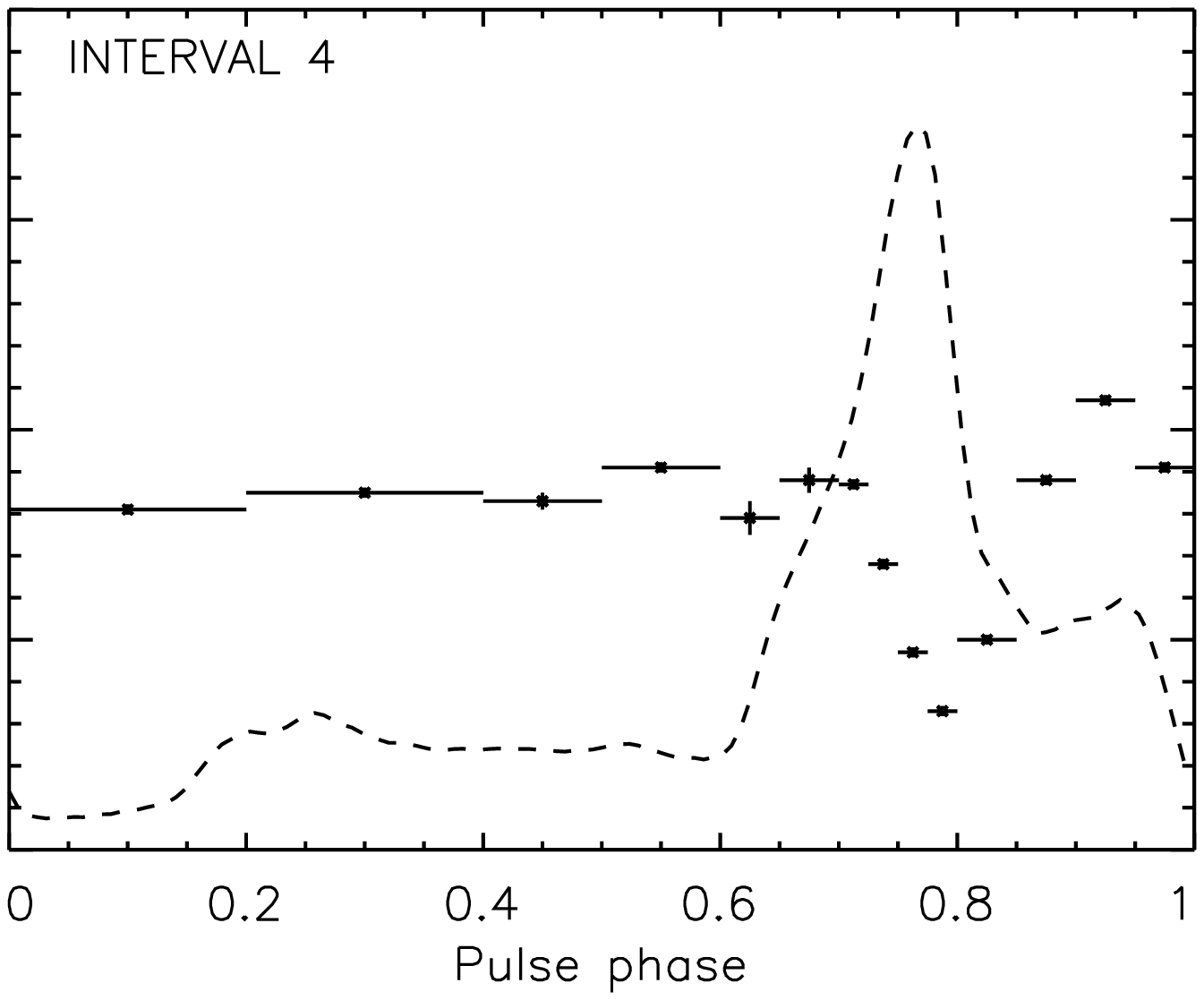}
   \caption[Photon-index profiles]{\small{Photon-index profiles as function of pulse phase for Interval 1(\textsl{top left}), Interval 2 (\textsl{top right}), Interval 3 (\textsl{bottom left}) and Interval 4 (\textsl{bottom right}).}}
   \label{pho}
\end{figure*}

%Fig. 7
\begin{figure}[hb!]
    \includegraphics[width=0.50\textwidth]{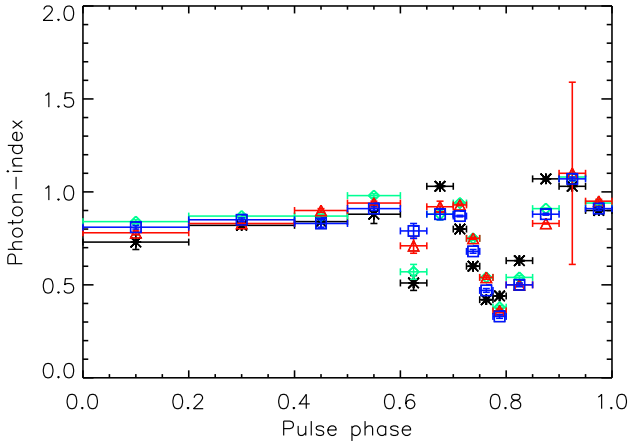}
   \caption{The photon index $\Gamma$ as function of pulse phase for the four different 35\,d phase
   intervals (see Table \ref{observations}):
  Interval 1 in black (asterisks), Interval 2 in green (diamonds),
  Interval 3 in red (triangles) and Interval 4 in blue (squares).}
   \label{photonindex6}
\end{figure}

\subsection{Cyclotron line energy}

The profiles of the cyclotron line energy E$_{\rm cyc}$ as a function of pulse phase are shown in 
Figures \ref{cyclll} and \ref{compar_e}.  Generally, E$_{\rm cyc}$ roughly follows the shape of the pulse 
profile: the broad maximum is found to be close to the peak of the pulse profile (around pulse phase 0.7) 
for all 35\,d phases. The formal uncertainties of E$_{\rm cyc}$ (as determined by XSPEC) are of the order 
of a few to several percent for pulse phases 0.0--0.6 and 0.9--1.0 and around one percent or less for pulse 
phases 0.6--0.9 (close to the peak of the pulse). 
Adopting an additional systematic uncertainty of 0.15\,keV ($\sim0.4$\%) for all data points,
the profiles are well represented by cosine functions (see below). 
We find slightly different values for the mean E$_{\rm cyc}$ and the peak-to-peak amplitude:
a slight, but significant increase is found as function of 35\,d phase.  The weighted mean values of the 
four highest values of E$_{\rm cyc}$ of each group are $40.16\pm0.22$\,keV, $40.89\pm0.20$\,keV, 
$41.02\pm0.15$\,keV and $41.52\pm0.20$\,keV for mean 35\,d phases 0.03, 0.10, 0.15 and 0.20 
(intervals 1 through 4), respectively. This corresponds to an increase by $\sim0.7$\,keV per 0.1 units in 
35\,d phase. The minimum E$_{\rm cyc}$ values have larger uncertainties with a tendency to decrease with
35\,d phase, but, within uncertainties, they are consistent with a constant value (around 33.3\,keV). 
The peak-to-peak amplitude is then slightly increasing with 35\,d phase (with a mean of $\sim7.6$\,keV).

%Fig. 8
\begin{figure*}[t!]
\centering
     \includegraphics[bb=95 417 533 694,width=8cm]{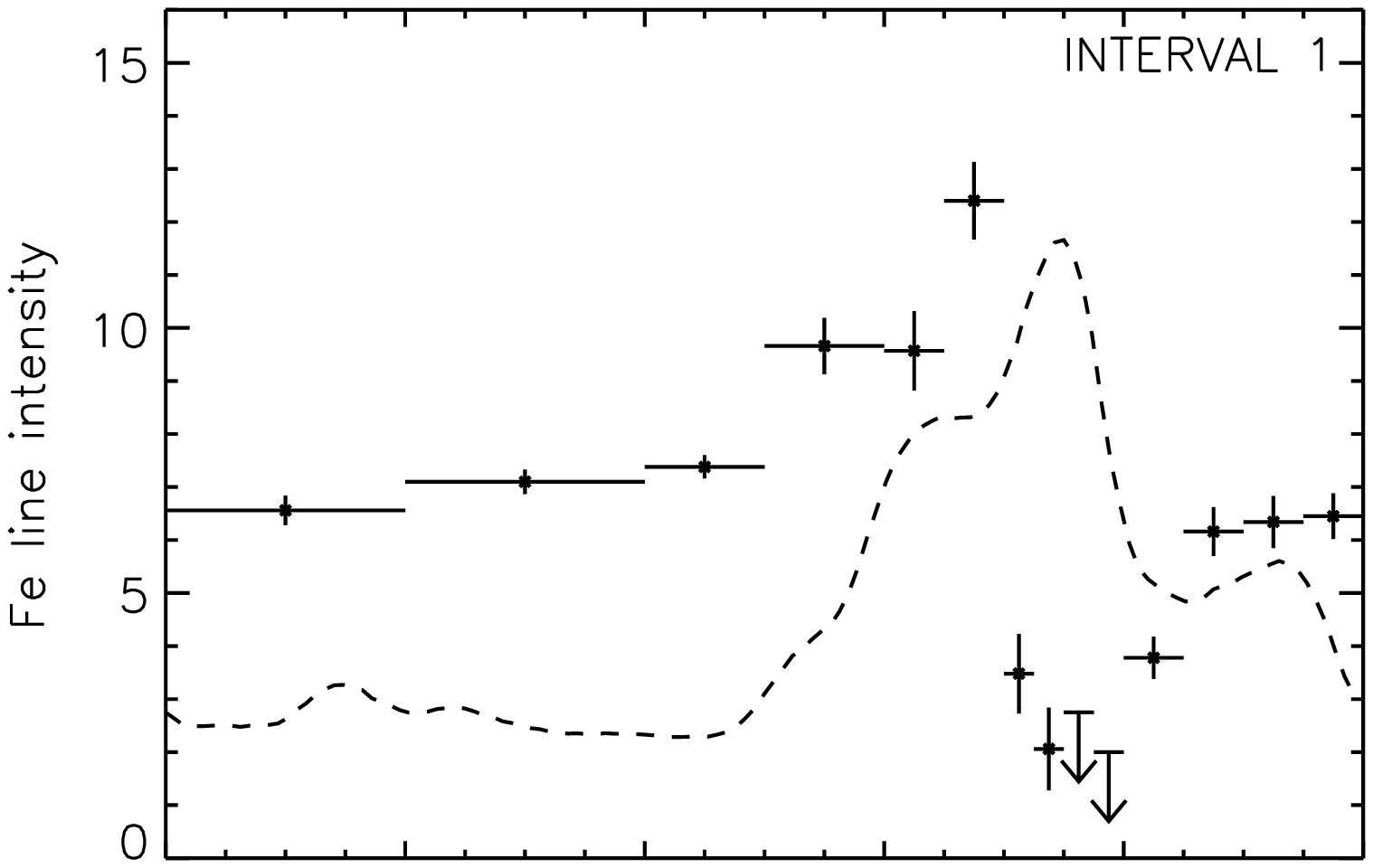} 
     \includegraphics[bb=143 417 533 694,width=7.125cm]{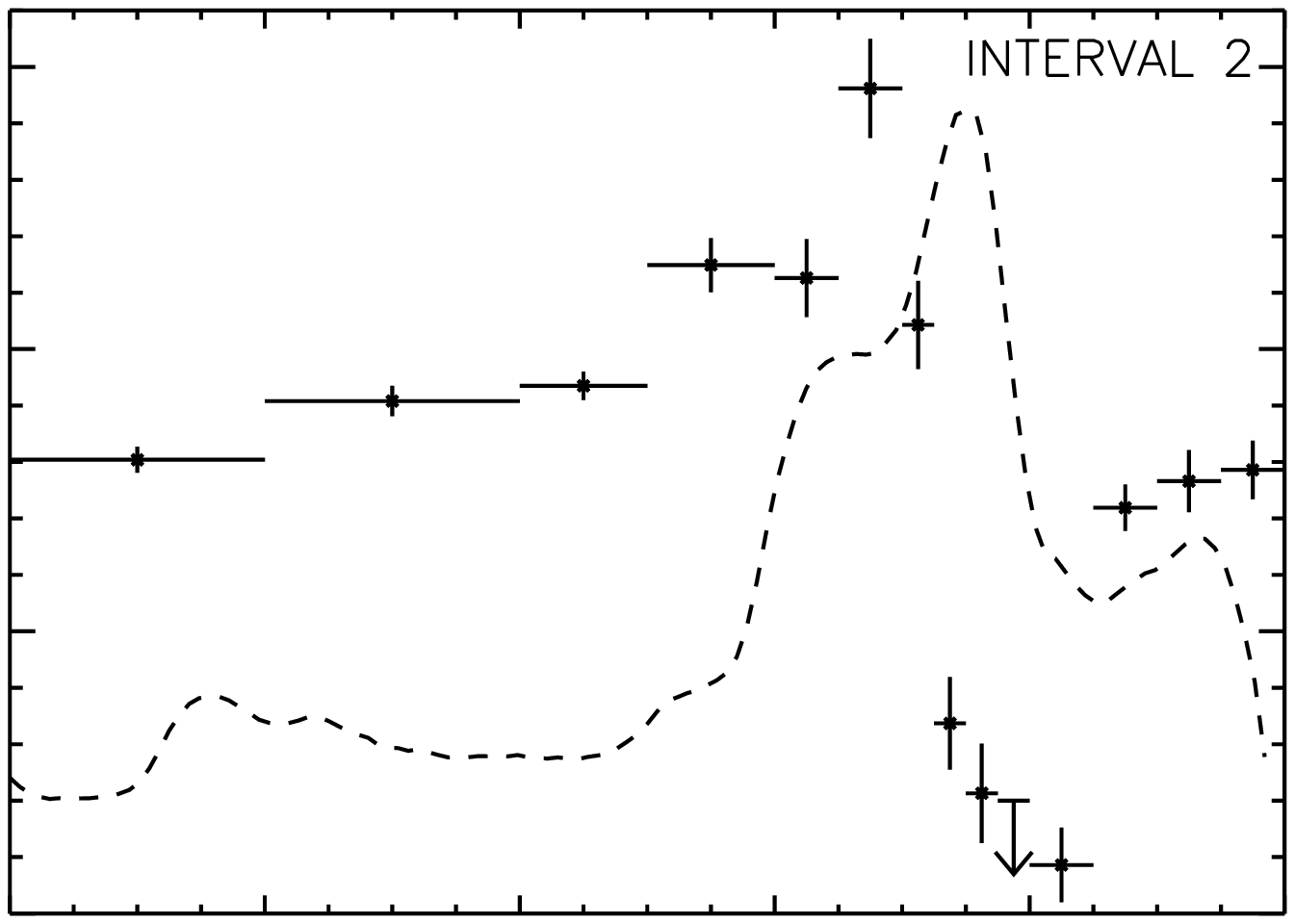}\\
     \includegraphics[bb=95 371 533 694,width=8cm]{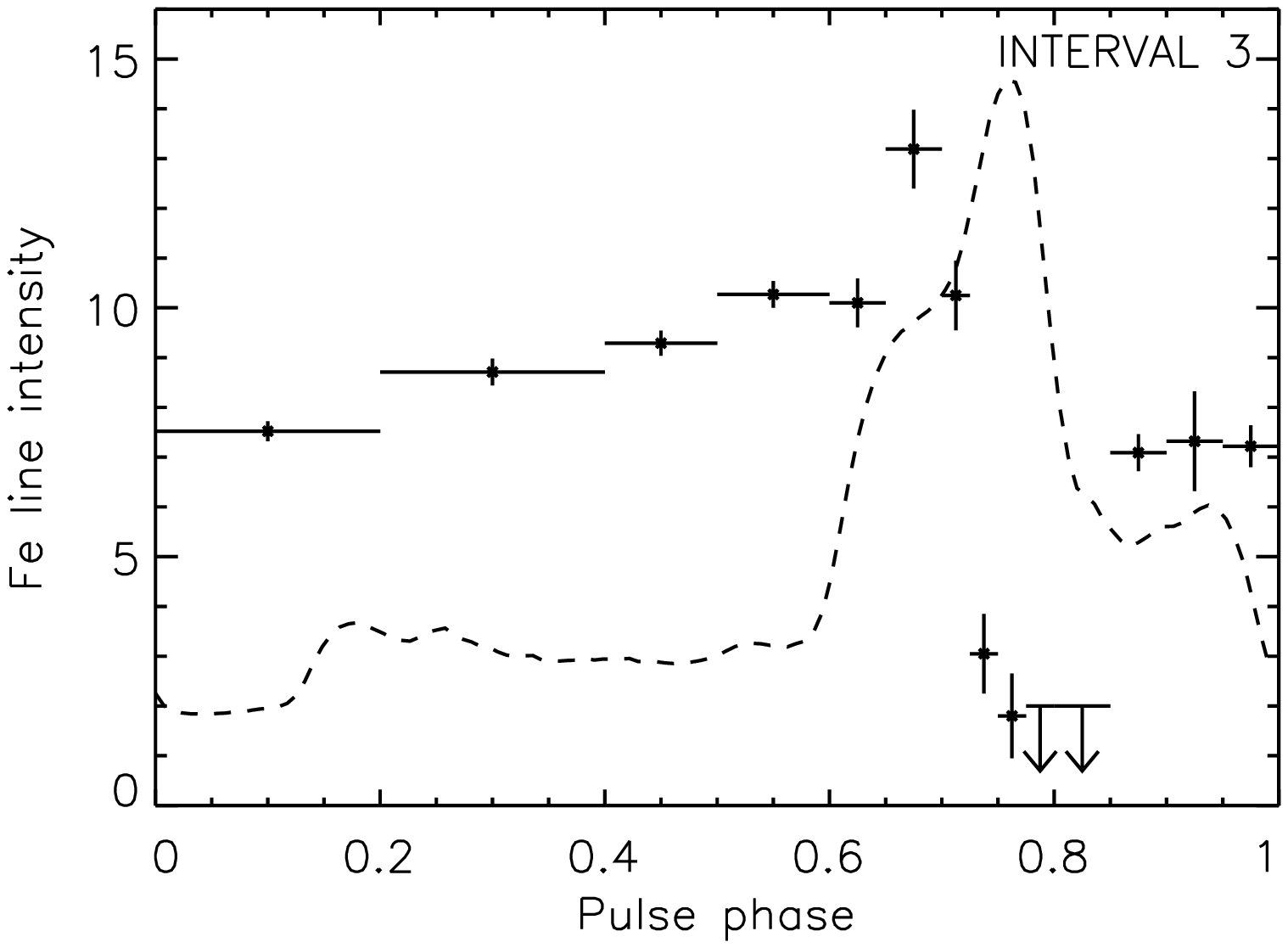}
     \includegraphics[bb=143 371 533 694,clip,width=7.125cm]{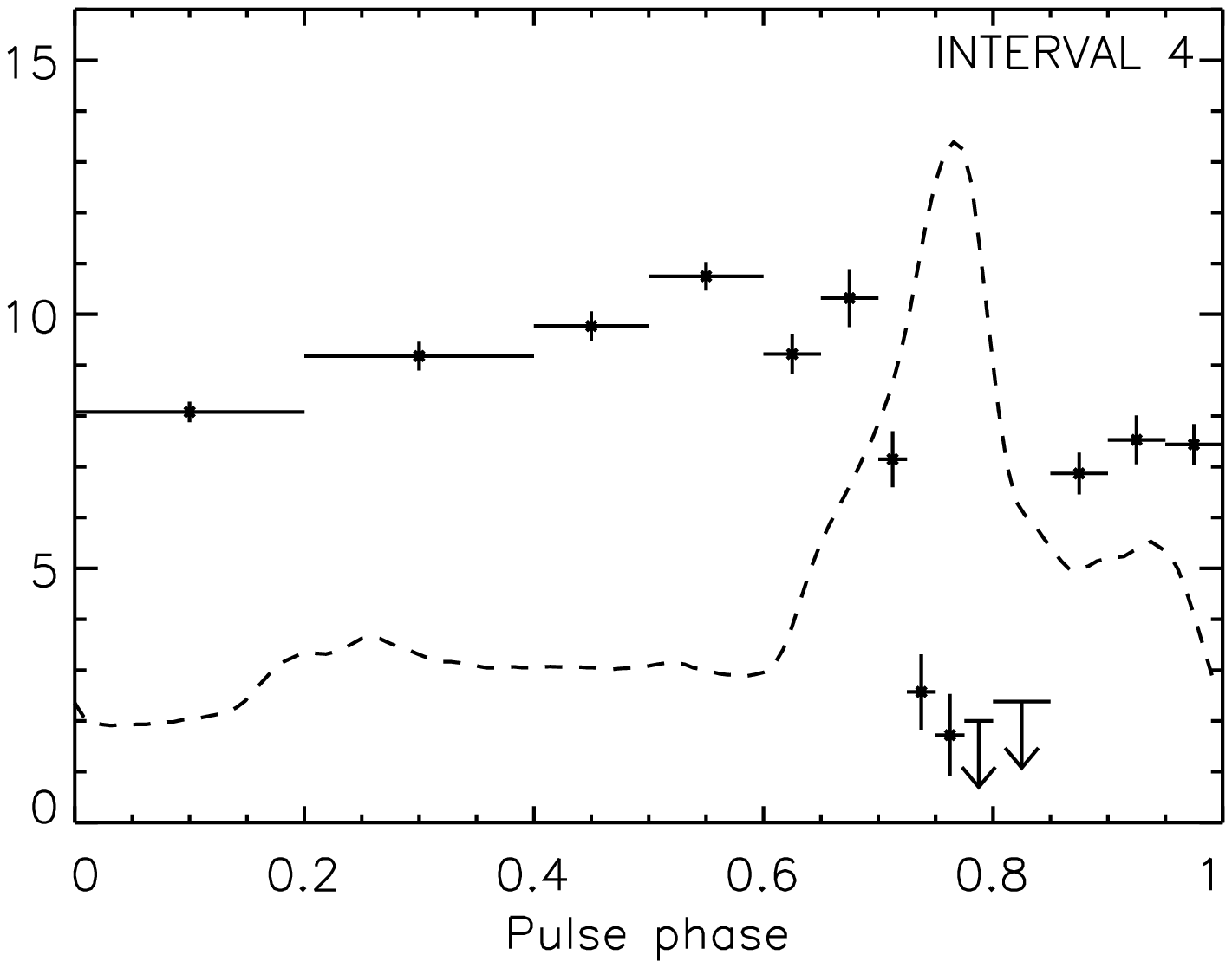}
   \caption[Iron line intensity profiles]{\small{Iron line intensity
       profiles as function of pulse phase for Interval 1(\textsl{top
         left}), Interval 2 (\textsl{top right}), Interval 3
       (\textsl{bottom left}) and Interval 4 (\textsl{bottom
         right}). All intensities less than 1.0 are consistent with
       zero, we therefore show two sigma upperlimits.}}
   \label{iron}
\end{figure*}

While the values for the maximum E$_{\rm cyc}$ and the peak-to-peak amplitude are dependent
on 35\,d phase, the sinusoidal shape of the modulation does not change. 
In order to verify this statement, the following analysis was performed: the original E$_{\rm cyc}$ 
profiles (Figure \ref{compar_e}) were scaled such that the separation of minimum to maximum 
E$_{\rm cyc}$ had a constant value of 7.64\,keV, using the minimum values measured in phase bin 
0.2--0.4 and the weighted mean of the four highest values in each of the four profiles as reference. 
In this way, \textsl{normalized E$_{\rm cyc}$ profiles} were generated.
These normalized profiles are shown in Fig. \ref{normalized2}, each in comparison to the weighted mean values 
and a smooth best fit function of the weighted mean: a combination of two cosine functions with common 
mean (37.31\,keV), amplitude (3.82\,keV), period and phase-zero values of 0.70 and 0.772 for pulse 
phases 0.3--0.875 and 1.47 and 0.3 for the remaining phases, respectively.
Visually, there is no obvious deviation of any of the four data sets from the weighted mean.
As quantitative statistical test we performed the following: comparing the individual normalized 
profiles with that profile representing the weighted mean leads to $\chi^{2}$ values (normalized 
to the degrees of freedom) of 0.72, 1.46, 0.72 and 1.32 for the four intervals, respectively. 
As an additional test, we have then
performed a Monte Carlo simulation, in order to answer the following question. Assuming the 
profiles are indeed the same for all four intervals, that is equal to the best fit mean profile: what is the
probability to find an individual profile which deviates from the mean profile in such a way that the
$\chi^{2}$ reaches or exceeds the observed value? We have simulated 100 profiles for each of the
four intervals, reproducing the statistical conditions of each
interval (that is, the width of the Gaussian
distribution from which the random numbers were drawn were chosen according to the uncertainty 
of the corresponding measured data point). 
Those probabilities are 68\%, 29\%, 73\% and 4\% for the four intervals, respectively. 
Even 4\% for Interval 4 is a rather high and acceptable probability.
We conclude, that, within our statistical accuracy, we find no evidence for any change in the shape of the 
E$_{\rm cyc}$ profiles with 35\,d phase.

%Fig. 9
\begin{figure}[b!]
    \includegraphics[width=0.50\textwidth]{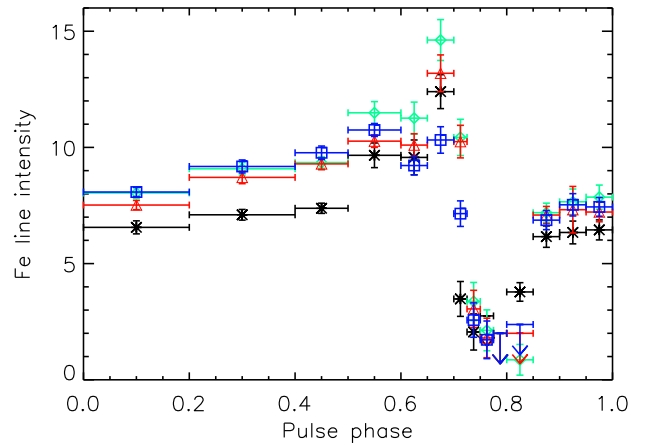}
   \caption{The iron line intensity profiles for the four different 35\,d phase intervals (see Table 
   \ref{observations}): Interval 1 in black (asterisks), Interval 2 in green (diamonds),
  Interval 3 in red (triangles) and Interval 4 in blue (squares).}
   \label{compar_i}
\end{figure}

\subsection{Photon-index}

Figure \ref{pho}  shows the profiles of the photon index $\Gamma$ as function of the pulse phase. This 
spectral parameter is relatively constant for pulse phases up to 0.6 (showing a small positive slope),
but has a distinct dip around the pulse peak and a smaller one around the left shoulder. The 
photon-indeces range between 0.42 $\pm$ 0.01 and 1.07 $\pm$ 0.01 for Interval 1, between 
0.54\,$\pm$\,0.01 and 1.08\,$\pm$\,0.02 for Interval 2, between 0.36\,$\pm$\,0.01 and 
1.10\,$\pm$\,0.49 for Interval 3 and between 0.33\,$\pm$\,0.01 and 1.07\,$\pm$\,0.01 for Interval 4. 
In general, the spectrum gets harder when the flux increases. 
While the main features of the pulse phase dependence of $\Gamma$ are very much the same 
in the four 35\,d intervals, the depth of the new narrow dip around phase 0.62 shows a clear trend
towards smaller values from Interval 1 to Interval 4 (see Fig. \ref{photonindex6}).

%Fig. 10
\begin{figure*}[t!]
\centering
    \includegraphics[bb=100 407 535 694,clip ,width=7cm]{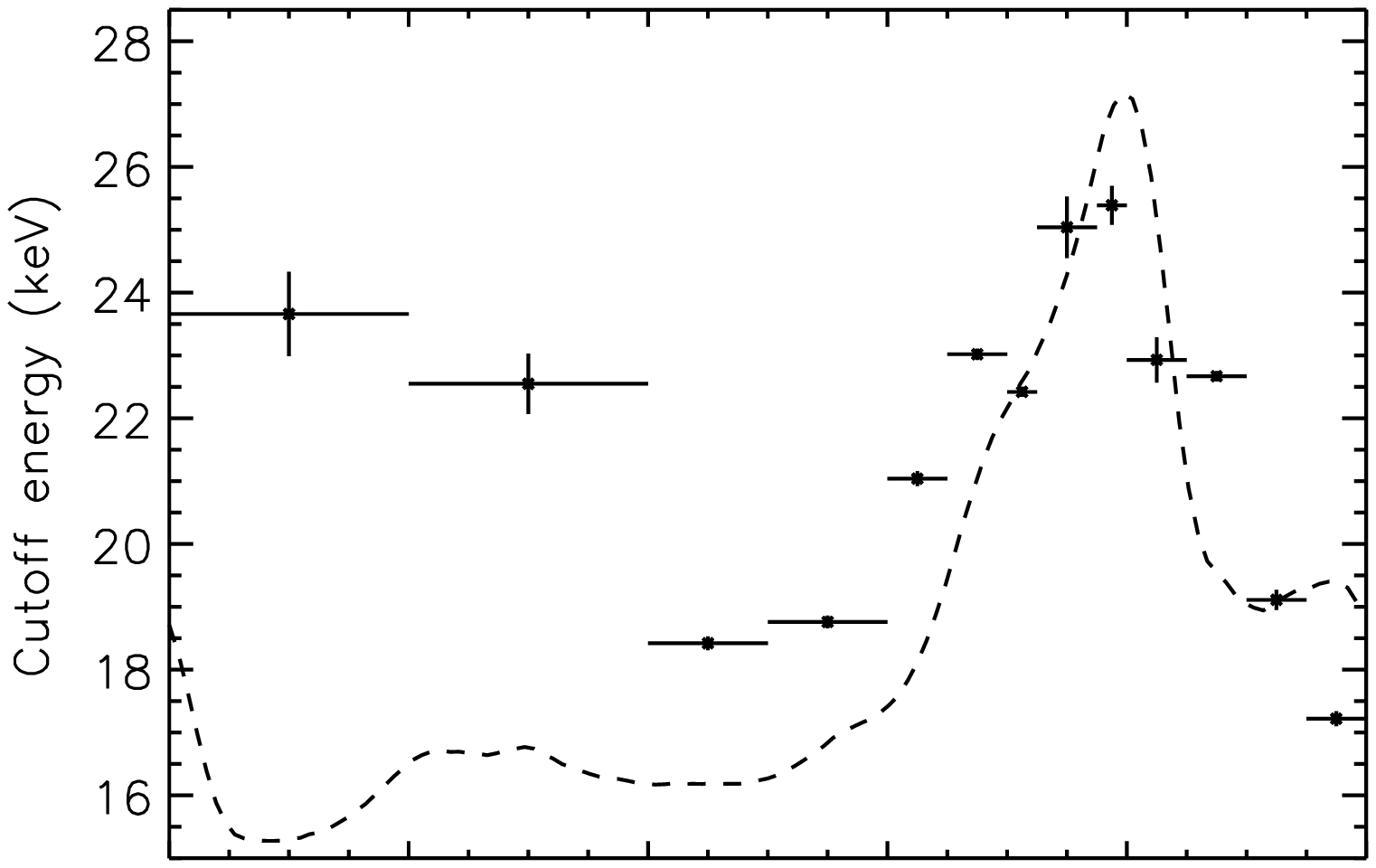} 
     \includegraphics[bb=100 407 535 694,clip,width=7cm]{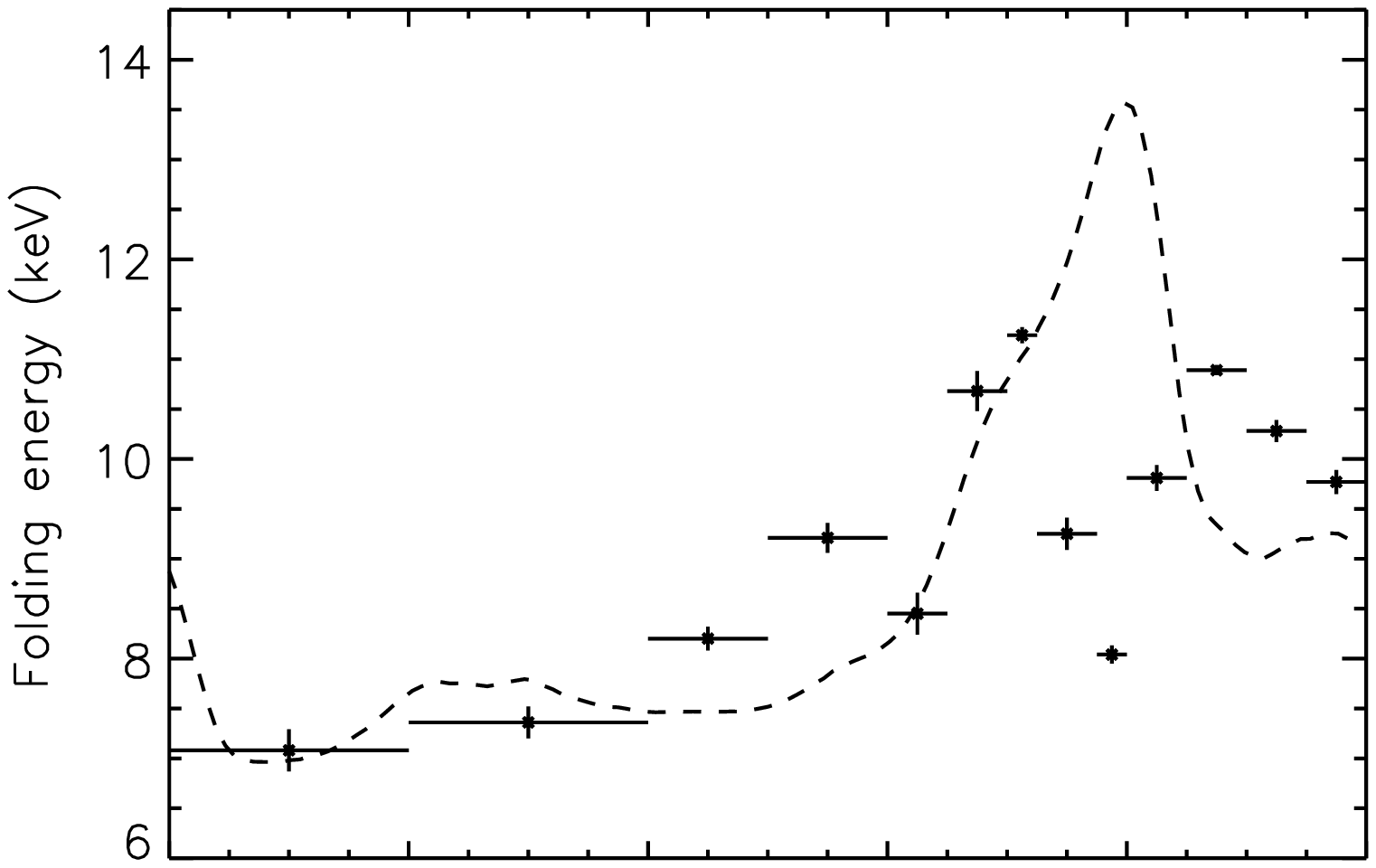}\\
   \includegraphics[bb=95 371 533 694,clip,width=7cm]{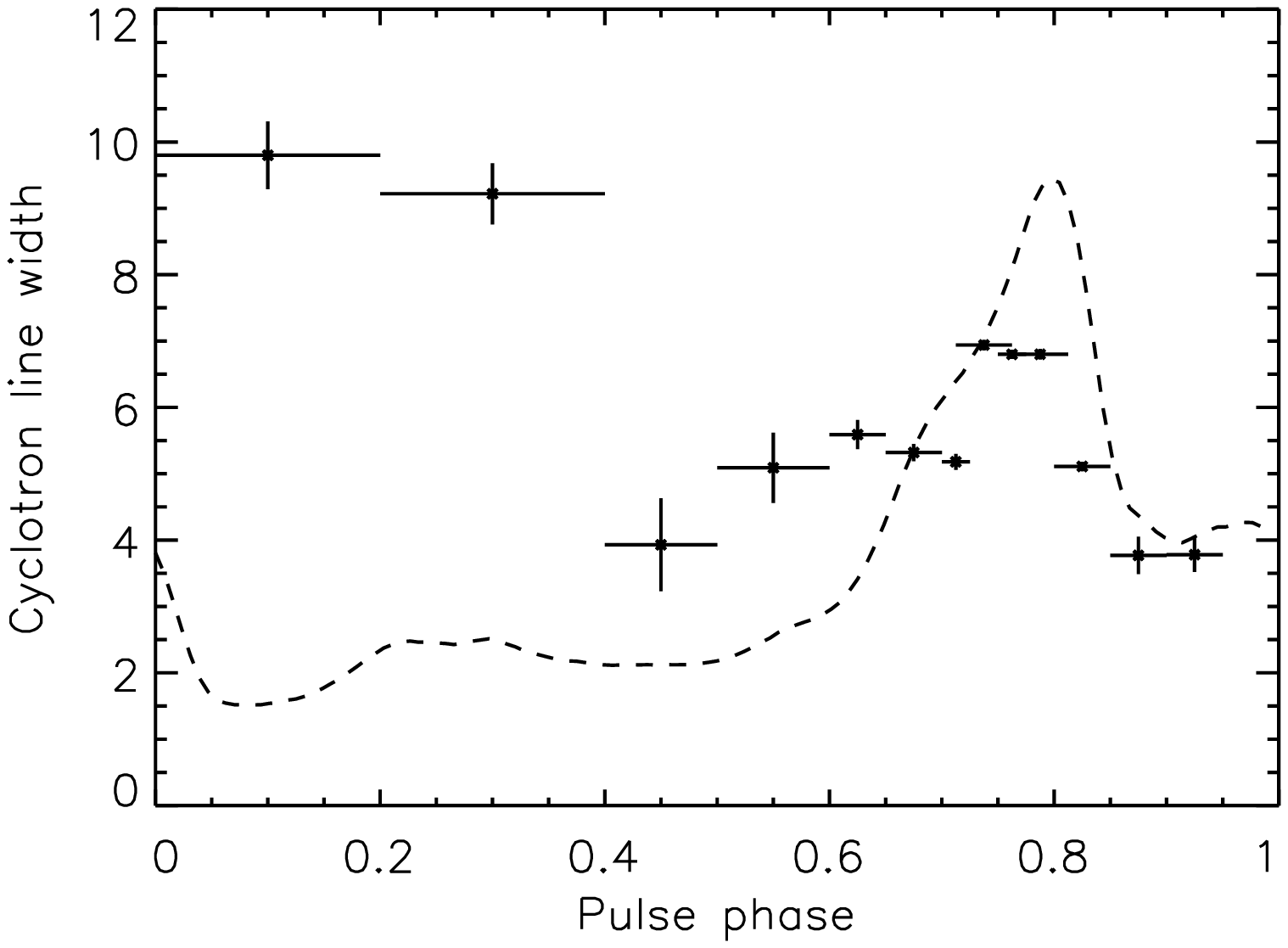}
     \includegraphics[bb=95 371 533 694,clip,width=7cm]{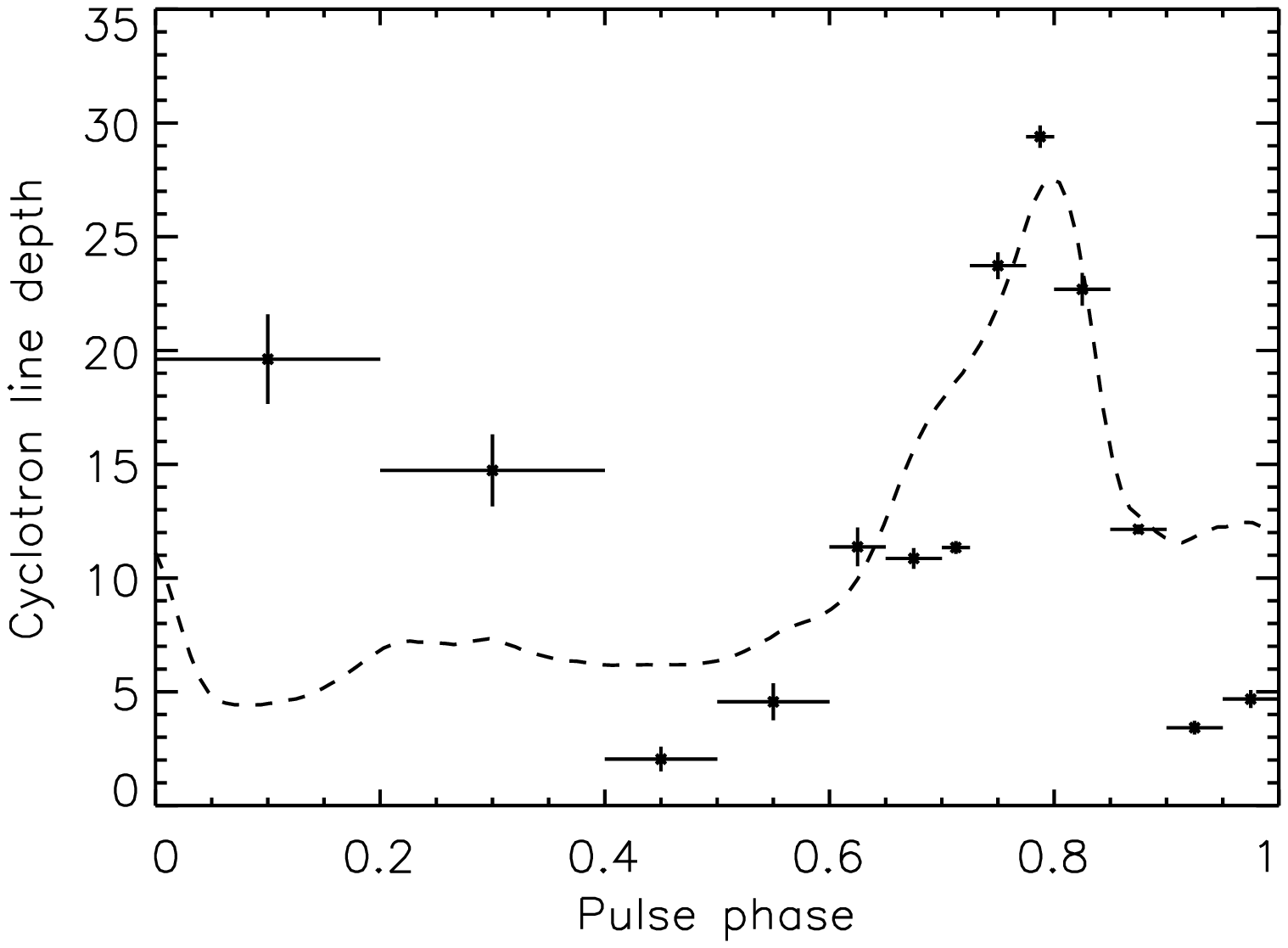}
   \caption{\small{Pulse phase profiles for the cutoff energy E$_{\rm cut}$ 
   (\textsl{top left}), the folding energy E$_{\rm fold}$ (\textsl{top right}), the width of the cyclotron line 
   $\sigma_{\rm cyc}$ (\textsl{bottom left}), and the depth  $\tau_{\rm cyc}$ of the cyclotron line 
   (\textsl{bottom right}) for the total data sets.}}
   \label{extra}
\end{figure*}

\subsection{Iron line intensity}

Another spectral parameter which shows a clear modulation as function of pulse phase
is the iron line intensity (see Figure \ref{iron}).
It shows a moderate increase until the left shoulder of the pulse (phase $\sim0.7$), after which it drops
almost to zero around pulse phase 0.8, before reaching the start value again at pulse phase 0.9. 
The minimum is coincident with the peak of the X-ray pulse (see Figure \ref{iron} and \ref{compar_i}) in a 
similar way as the photon index. The iron line intensity ranges between 0 (for all intervals) and 12.40 for 
Interval 1, 14.62 for Interval 2, 13.19 for Interval 3 and 10.75 for Interval 4. The profiles of 35\,d 
intervals 2--4 are quite similar to one another, while for Interval 1, the iron line intensity is  
significantly lower for pulse phases $<$ 0.6.

\subsection{Other spectral parameters}

For the other spectral parameters E$_{\rm cut}$, E$_{\rm fold}$ the width $\sigma_{\rm cyc}$ and the depth 
$\tau_{\rm cyc}$,  we show their pulse phase dependence only for the total data sets (not resolved in 35\,d 
phase), because in these parameters we find a considerably larger scatter, such that no conclusions can 
be drawn about a possible variation (or not) with 35\,d phase  (the 35\,d phase resolved profiles can
be found in \citealt{Vasco_12})\footnote{http://tobias-lib.uni-tuebingen.de/frontdoor.php?source$\_$opus=6346}. 
The profiles are shown in Figures \ref{extra}. Also for these parameters there is significant modulation with 
pulse phase, and they generally show a broad maximum around the phase
of the pulse peak, with the
interesting feature that for E$_{\rm fold}$ there is a dip in this maximum, right at the pulse peak.
It is also worth noting, that, except for E$_{\rm fold}$, the measured values for pulse phases $<$ 0.4 are
rather high, not following the shape of the pulse profiles.

\subsection{The "10 keV feature"}

It has been noticed (e.g. \citealt{Coburn02}), that in fitting observed spectra of accreting X-ray binaries 
with a cut-off power law a "wiggle" or "bump" around $\sim10$\,keV to $\sim15$\,keV is found in the 
residuals. This applies to many different sources including Her~X-1 (also to non-cyclotron line 
sources), to observations with different satellites (e.g. Ginga,
\emph{Beppo}/SAX and RXTE), 
and to analysis with different spectral functions involving a power law (e.g., \texttt{highecut,plcut}, Fermi-Dirac 
or NPEX). The feature is therefore considered to be inherent to accreting X-ray binaries. So far, no
physical explanation has been suggested.

\begin{figure}[bh!]
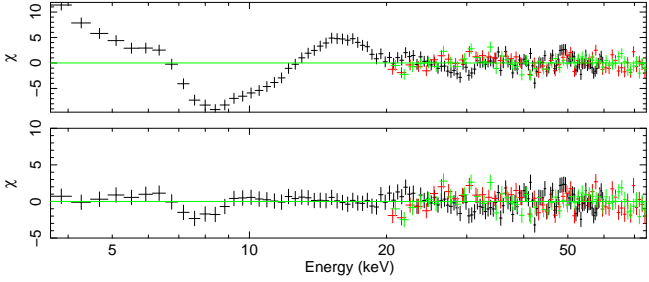

     \centering
     \includegraphics[bb=413 45 534 701,clip,angle=-90,width=8.5cm]{res-1.ps}
 \includegraphics[bb=408 45 570 701,clip,angle=-90,width=8.5cm]{res2-1.ps}
   \caption{Residuals of the X-ray spectrum of pulse phase bin
     0.80\,-\,0.85 for Interval 3
    without (top) and with (bottom) an extra Gaussian component in the fit.}
   \label{res}
\end{figure}

In our current analysis, using data from observations of Her~X-1 by RXTE, this feature does not play a 
role in the phase averaged analysis. 
However, in the pulse phase resolved analysis, larger $\chi^{2}$ values appear due to the existence 
of this "10 keV feature": up to $\sim$12 for the pulse phase range 0.775--0.85, and 1.4--2.8 for the pulse 
phase range 0.5--0.775. Modeling this feature by an extra Gaussian is generally successful and brings the 
$\chi^{2}$ down to acceptable values (in a few cases $\chi^{2}$ is still found around 1.3, which is most likely 
due to an imperfect modeling by  a simple Gaussian).
The centroid energy is generally found at $\sim16$\,keV with a $\sigma$ of $\sim5$\,keV.
Fig. \ref{res} shows an example of the spectrum of Interval 3, pulse phase bin 0.80--0.85, where the
$\chi^{2}$ is reduced from 5.55 to 1.38 for 224 dof.
We have verified, that all the other spectral parameters are not altered when
introducing this extra \emph{Gaussian}. We note that this feature occurs most strongly at a pulse phase 
close to the peak of the pulse, where all other spectral parameters also tend to show their strongest 
variability.

\section{Summary and discussion}

We have selected the best data available in the RXTE archive of an observation
of a Main-On of Her~X-1 to perform detailed pulse phase resolved spectroscopy
in the energy range 3.5--75\,keV. The observations  of November 2002 (cycle no. 323)
provide the best RXTE coverage of a Main-On \citep{Vasco_12}, such that an 
attempt could be made to study pulse phase dependent spectral variations also as a function 
of the phase of the 35\,d  modulations - flux and shape of the pulse profiles -, which are believed 
to be due to precessional motion of the accretion disk and, possibly, the neutron star itself.  
As spectral model we have used a power law 
with an exponential cut-off (the \texttt{highecut} model in XSPEC) modified by two line features, 
a multiplicative cyclotron line (in absorption) and an additive iron fluorescence line (in emission). 
We have concentrated on three spectral parameters: the centroid energy  of the 
cyclotron line, the photon index $\Gamma$ of the power law component (valid between 3\,keV and 
$\sim20$\,keV), and the intensity of the iron fluorescence line at 6.4\,keV. For these three parameters, 
the analysis was performed for four one day long intervals at different mean 35\,d phases (0.03, 0.10, 
0.15 and 0.20,  respectively). For the other spectral parameters of the applied spectral model, no 
35\,day phase dependence was studied. 

%Fig. 11

\subsection{Pulse phase dependence}

The results of the pulse phase resolved analysis can be summarized as follows.
We distinguish between results which \textsl{confirm and improve} in statistical
significance which was known before and \textsl{new} results.

\subsubsection{Confirmed and improved results}

\begin{itemize}
\item All spectral parameters show a strong modulation with pulse phase.
\item The strongest variations always appear around the peak of the pulse (at maximum flux).
\item The profile of the centroid energy E$_{\rm cyc}$ of the cyclotron line follows roughly the 
modulation of the pulse profile (maximum energy around maximum flux). 
The shape is close to "sinusoidal".
\item The photon index profile (Fig. \ref{pho} and \ref{photonindex6}) shows a strong dip close to the pulse peak. The dip around
the main peak is a known feature, nicely confirmed here and very well resolved in this analysis. 
\end{itemize}

\subsubsection{New results}

\begin{itemize}
\item It is the first time that for Her~X-1 a phase profile of the centroid cyclotron line energy 
E$_{\rm cyc}$ is produced with such a high resolution and statistical accuracy.
The four smallest bins around the peak of the pulse have a width of 1/80 of a phase.
The mean variation (minimum to maximum) is $\sim23$\% (Figs. \ref{cyclll} and \ref{compar_e}).
The modulation is close to "sinusoidal"; however, at least two sinusoidal components (with
greatly different periods) are needed to model the modulation.
\item For 35\,d intervals 1--3, the peak of the E$_{\rm cyc}$ profile \textbf{is} structured (see Fig. 
\ref{cyclll}), with local maxima near the maximum of the main peak of the pulse profile and around the
position of the left shoulder of the main peak. A similar structure is seen in $\tau_{\rm cyc}$.
\item The folding energy E$_{\rm fold}$ follows the pulse profile, except for a narrow dip in the peak of
the pulse (Fig. \ref{extra}, upper-right).
\item Similarly, the cut-off energy E$_{\rm cut}$, the width and depth of the cyclotron line ($\sigma_{\rm cyc}$ 
and $\tau_{\rm cyc}$), roughly follow the pulse profile, except for higher values at phases 0.0--0.5 
(Fig. \ref{extra}).
\item While the strong dip in the photon index profile
    (Fig. \ref{pho} and \ref{photonindex6}) close to the main pulse peak was
known (even though less well measured), the second, very narrow dip (around phase 0.61) is a newly 
detected feature. We note that this is close to the position of the left shoulder component of
the main peak.
\item The profile of the intensity of the 6.4\,keV iron line show a distinct minimum at maximum flux 
(Figs. \ref{iron}, and \ref{compar_i}).
\item We find that the so called "10 keV feature" seen in the spectra of several accreting binary
pulsars, is around 16\,keV in Her~X-1, with a strong pulse phase dependence: essentially 
concentrated in the narrow pulse phase interval 0.8--0.9. Spectral fits for all the other phase 
intervals do not show significant residuals.
\end{itemize}

\subsection{Precession phase dependence}

With respect to any variation of the spectral parameters (or their pulse phase variations) with 35\,d phase, 
we consider only the cyclotron line energy E$_{\rm cyc}$, the photon index $\Gamma$ and the
intensity of the iron fluorescence line at 6.4\,keV. For the other parameters the photon statistics
is not sufficient to distinguish between different 35\,d phases. It is the first time that a study
of the variation of spectral parameters with 35\,d phase was done. So, the following results
are all \textsl{new}.
\begin{itemize}
\item The maximum value for E$_{\rm cyc}$ (reached near the pulse peak) shows a slight, but 
significant increase with 35\,d phase. The mean value is $41.0\pm0.1$\,keV, and the slope is
$\sim0.7$\,keV per 0.1 units of 35\,d phase.
\item The minimum value for E$_{\rm cyc}$ (around pulse phase 0.3) is formally consistent with
a constant value of $33.5\pm0.5$\,keV (we would not rule out a small decrease). Together with the 
maximum increasing, this means that also the peak-to-peak amplitude (mean value $7.6\pm0.5$\,keV) 
is increasing with 35\,d phase.
\item The shape of the E$_{\rm cyc}$ profile (see Figs. \ref{compar_e} and \ref{normalized2}) \textbf{is}
independent of 35\,d phase.
\item Similarly, the shape of the $\Gamma$ profiles is independent of 35\,d phase, except for the
new narrow dip around pulse phase 0.62 (Figs. \ref{pho} and \ref{photonindex6}), where a trend to
a smaller depth from Interval 1 to Interval 4 is seen.
\item The shape of the iron line intensity profiles are similar to one another, except for Interval 1
which shows a distinctly lower intensity for pulse phases $<$0.6 (Figs. \ref{iron} and \ref{compar_i}).

\end{itemize}

\subsection{Comparison with previous results and physical conclusions}

For Her X-1, the dependence of the spectral parameters on pulse phase have been 
studied by several authors on the basis of observations with different X-ray instruments
(see Introduction). In the following we highlight our new findings and significant improvements
and discuss possible physical implications.

\textsl{Pulse phase dependence of the centroid cyclotron line energy}: This modulation
is quite  common among accreting X-ray pulsars and is generally believed to be due to
the changing viewing angle under which the X-ray emitting regions are seen 
(e.g., \citealt{Kreykenbohm04} and references therein). If we adopt the idea, that peaks
(high X-ray flux) in the pulse profile can be associated with beamed radiation emitted from
the accreting regions around the magnetic poles, then the observed pulse profiles are
a representation of the emission characteristics. During a full neutron star rotation we may
be seeing different emitting spots (most likely from a complex magnetic field configuration
which deviates from a simple dipole (see e.g. \citealt{Suchy08}), possibly having a multi-pole configuration)
and the emerging beamed radiation under changing angles. The situation is further complicated
by gravitational bending. It is therefore not straightforward to interpret observed
pulse profiles in terms of accretion geometry and beaming characteristics (see e.g. \citealt{Caballero11}).
If we, however, assume that the main peak in the pulse profile of Her~X-1 is due to a
reasonably narrow pencil beam (following e.g., \citealt{Pravdo78, Klochkov08}), emitted 
perpendicular to the neutron star surface at the hot spot of the main accreting pole, then 
we may associate this pulse phase with a situation where the observer is looking down 
into the accretion column, parallel to the magnetic field lines and
the velocity of the
in-falling material. At this phase we see the maximum cyclotron line energy
and the hardest spectrum (a sharp minimum in photon index $\Gamma$) - that is, we are 
looking deep down to small distances from the neutron star surface where the magnetic field 
is strongest and the temperature is highest. Here, we also find the largest width and depth 
of the cyclotron line, as well as the highest E$_{\rm cut}$ and a local dip within a broad peak of 
E$_{\rm fold}$.

Along similar lines, the newly discovered structure of a second (very sharp) minimum
in $\Gamma$ around pulse phase 0.61 (Fig. \ref{photonindex6}), corresponding to
the left shoulder of the main peak, may be the  signature of a second pole. At this
pulse phase we also find indications for another relative maximum in E$_{\rm cyc}$
(Fig. \ref{cyclll}). \\

\textsl{Pulse phase dependence of the 6.4\,keV iron line flux}:
Figs. \ref{iron} and \ref{compar_i} show a sharp and deep minimum of the intensity 
of the 6.4\,keV iron line, again coincident with the main pulse peak. The profile is very
similar to that of $\Gamma$ (Fig. \ref{photonindex6}) and the Fe line intensity drops to near 
zero at pulse phase $\sim\,$0.79 (where $\Gamma$ is minimal). It is the first time, that
such a sharp minimum in the 6.4\,keV flux is observed.  While \citet{Choi_etal94}, in 
observations by Ginga, and \citet{Zane04}, in observations by \emph{XMM-Newton}, had 
found a sinusoidal-like modulation with a broad minimum around the pulse peak,
\citet{Oosterbroek00}, with data from \emph{Beppo}/SAX, were only able to measure
the Fe-L~line around 1\,keV, for which they had found a minimum around the pulse peak
which was similarly sharp and deep as we now find for the 6.4\,keV line.
It is generally believed, that the origin for the Fe lines is fluorescence in relatively cool
material which is being illuminated by the X-ray beam. For the geometrical site, however,
where this fluorescence is taking place, or whether it is predominantly in reflection
or in transmission, no general consensus has been reached so far. Several different regions
have been proposed (see. e.g. \citealt{Choi_etal94} and references therein): 
1) the outer edge of the accretion disk extended in the vertical direction; 2) a corona of the 
accretion disk; 3) the atmosphere of the companion star; 4) a surface at the Alfv\'en radius.
The Alfv\'en surface is currently the most popular one. However, also the
accretion column was proposed (e.g., \citealt{Leahy01} and \citealt{Inam05}), despite the
argument by \citet{White83} that the material in the column may be too hot.

We like to stress the observed fact, that at the pulse phase where we see the highest 
continuum flux there is essentially no Fe line flux. This should provide some constraints in
finding a feasible geometry. If the above adopted scenario is correct, by which we look
down into the pencil beam at the pulse peak, the Fe line result may be consistent with a hollow
cone geometry, where the majority of the radiation escapes through the empty center of
the cone, where there is no material which can fluoresce, while the walls are seen at 
off-center angles. It is, of course, assumed that the bottom of the cone close to the 
neutron star surface is filled with X-ray emitting plasma.

\textsl{The non-dependence of the cyclotron line energy profile on 35\,d phase and the model 
of neutron star free precession}:
One of the results of this study is the finding that the shape of the cyclotron line energy 
profile (E$_{\rm cyc}$ vs pulse phase) \textbf{is} independent of 35\,d phase
(Figs. \ref{cyclll} and \ref{compar_e}), while the maximum cyclotron line energy 
increases slightly by $\sim$0.7\,keV per 0.1 units of 35\,d phase. If we attribute the
strong variations in pulse profile with 35\,d phase to precession of the neutron star, that
is to the change of the angle under which we see the X-ray emitting regions, we would
expect also the shape of the cyclotron line energy profile to change with 35\,d phase.
However, this expectation needs a quantitative analysis. Recently concluded detailed 
model calculations \citep{Postnov_etal12}, assuming neutron star free precession, 
have shown, that under a specific set of assumptions (including a certain multi-pole 
geometry and emission into narrow pencil beams), the observed pulse profiles in the 
energy range 9-13\,keV (the same as used in \citealt{Staubert_etal10a,Staubert_etal10b,
Staubert_etal12}) can be reproduced with high accuracy. 
We point out here that the model of \citet{Postnov_etal12} is an attempt to give
a physical explanation of the variation in pulse profiles by neutron star free precession,
requiring several specific assumptions, while the template approach of \citet{Staubert_etal12},
mentioned in the Introduction, does not need any physical assumptions, it just makes use
of the observed pulse profiles. We plan to investigate whether 
the here described constancy of the shape of the cyclotron energy profile is consistent 
with the physical model. In this sense our result opens a new channel to test the model of free 
precession of the neutron star.

 \begin{acknowledgements}
      This paper is based on observational data taken by the NASA satellite
      \textsl{Rossi X-ray Timing Explorer} (RXTE). We like to acknowledge the dedication 
      of all people who have contributed to the great success of this mission. 
      D.V. and coauthors thank DLR for financial support through grant
      50 OR 0702 and D.K. acknowledges support by the Carl Zeiss Stiftung.
      We thank the anonymous referee for very valuable comments.
 \end{acknowledgements}

\bibliographystyle{aa}
\bibliography{pprs_davide_18Dec12} 

 \end{document}